\PassOptionsToPackage{super}{natbib}%% AMA-style superscript numerals
\documentclass[pdflatex,sn-vancouver-num]{sn-jnl}

\usepackage[T1]{fontenc}%% REQUIRED: OT1 renders bare > as ¿ and < as ¡
\usepackage{xurl}%% allow long URLs to break anywhere (reference list)
\makeatletter
\def\bibfont{\reset@font\fontfamily{\rmdefault}\small\selectfont}
\makeatother
\AtBeginDocument{\setlength{\bibsep}{0.6em}}
\usepackage{graphicx}
\usepackage{booktabs}
\usepackage{tabularx}
\usepackage{array}
\usepackage{multirow}
\usepackage{tikz}
\usetikzlibrary{arrows.meta,positioning,fit,calc,backgrounds}
\usepackage{amsmath,amssymb}
\usepackage{enumitem}
\usepackage{ragged2e}
\usepackage{url}
\usepackage[section]{placeins}

\setlist[itemize]{leftmargin=1.2em,itemsep=0.25em,topsep=0.35em}
\setlist[enumerate]{leftmargin=1.4em,itemsep=0.25em,topsep=0.35em}

\newcommand{\fullmark}{\ensuremath{\bullet}}
\newcommand{\partmark}{\ensuremath{\oslash}}
\newcommand{\absnmark}{\ensuremath{\times}}

\begin{document}

\title[Persona--Execution Separation]{%
  \texorpdfstring{%
    \begin{tabular}{@{}c@{}}
    Persona--Execution Separation:\\
    An Architecture Pattern for Evolving LLM Agents\\
    under Execution Audit
    \end{tabular}}{Persona--Execution Separation: An Architecture Pattern for Evolving LLM Agents under Execution Audit}}

\author*[1]{\fnm{Yisen} \sur{Xi}}\email{xys21@tsinghua.org.cn}
\affil*[1]{\orgname{Independent Researcher}, \orgaddress{\city{Beijing}, \country{China}}}

\abstract{Large language model (LLM) agents in governed organizations must let the persona (instructions, tone, self-presentation) evolve freely, while keeping execution (stateful, audited work) traceable. A single trust domain does not satisfy both cheaply. We present \textbf{Persona--Execution Separation (PES)}: persona and execution reside in different trust domains, connected by a governed contract bridge. The persona is singly-homed and may drift; execution is faceless and audited. Status summaries may return; data bodies remain in the restrictive domain except a data-loss-prevention (DLP) exception; identity stays continuous. An approval matrix, DLP, and audit enforce the crossing. PES follows from three goals: free drift, execution traceability, and decoupling. Under LLM representational indistinguishability, any single-domain mechanism meeting all three must re-introduce typed change objects, an external gate, and a stable audit anchor: PES rebuilt at higher coupling cost. A development/pilot case in a regulated platform records five decisions over one month, four with rejected alternatives. A mechanism check found no execution-side re-validation under persona perturbation (five configurations) and no persona fingerprint on hard-asserted fields of completed runs. A controlled replication in regulated coding agents reproduced the separation under isolation across five models and four providers; bridge overhead was under 0.2\% of end-to-end time in both environments. A probe of a pre-separation build found the execution path decoupled from the persona by omission, not by construction. The pattern applies when multi-user deployment, execution audit, and persona churn hold jointly.
}


\maketitle

\section{Introduction}

Large language model (LLM) agents now execute state-changing actions on corporate assets (fund operations, compliance filings, due-diligence work) and inherit a demand that chat assistants never faced: \textbf{execution traceability}. Every state-changing action must be gated, recorded, and auditable. The same agents are also expected to \textbf{evolve}: operators continuously tune instructions, tone, and skill bindings, because that is how the system improves. Conventional agent architectures resolve the conflict by sacrificing one of the two.

In a single-domain agent --- the default design, where the agent's persona (its identity, instructions, presentation) and its execution (its actions on organizational assets) share one process and one trust domain --- the conflict is structural. If the domain is governed strictly, every persona edit triggers re-validation of everything the agent does. If it is governed permissively, execution is no longer reliably traced. A single governance regime does not serve both cheaply when the two concerns are representationally indistinguishable (Section 3.2). In practice organizations then freeze the persona and stall evolution, or loosen execution governance and lose a reliable audit trail.

An agent's persona and its execution need not reside in the same trust domain. \textbf{Persona--Execution Separation (PES)} puts the expression surface (where the agent is seen, talked to, and edited) and the execution surface (where it acts and is audited) in different trust domains, connected by a governed contract bridge (Section 4). The pattern is motivated by free persona drift (G1), strict execution traceability (G2), and their decoupling (G3).

Once agents take on regulated work, coupling persona evolution with execution traceability is an architectural problem, not a prompt-tuning one. Healthcare and public administration illustrate the \textit{same tension}; they are not additional cases in this paper (Section 8.1). Regulated coding agents illustrate it too, and carry a controlled replication of the mechanism claims (Section 7.5). Open-source platforms either couple the concerns in one domain, externalize execution as stateless tools (losing employee-level trace), or place governance as an external layer around execution rather than around the persona--execution relationship. Practitioners report governance and reliability challenges with agent frameworks \cite{hamid}; recent work separates \textit{intent} from execution for security, and hardens agent behavior into workflows over time. None of that work separates the \textit{persona} from execution so that the persona can keep changing.

\textbf{Contributions.} This paper makes three contributions.

\begin{enumerate}
\item \textbf{An architecture pattern \cite{posa}.} PES places an agent's operational identity and its governed execution in different trust domains, with a contract bridge that keeps one employee identity continuous. Execution semantics stay unchanged under persona edits if the bridge is enforced (Section 4.1). The pieces have precedents (Section 2.3); the combination does not, on a 2022--2026 scan (Section 6). We do not claim a formal guarantee.
\item \textbf{A development/pilot case with a decision chain.} We document how PES emerged in a digital-employee platform for financial institutions, five decisions over one month (persona storage, capability binding, one-way valve, promotion channel, dual-face crystallization), four with recorded rejected alternatives (Section 5; the seed decision predates the habit). A probe of a recovered pre-separation build found the governed execution path decoupled from the persona by omission (Section 7.4). A controlled replication in a second domain (regulated coding agents, on a standalone harness) reproduced the separation under isolation across five models and four providers and measured bridge overhead in two environments (Section 7.5). The platform is a development/pilot deployment, not a production system.
\item \textbf{A comparison with existing platforms.} Systematic comparison with eight open-source platforms and the academic neighbors of Section 6, including the 2025--2026 secure-execution cluster, shows no existing architecture jointly satisfies G1--G3. Section 6 operationalizes those goals as three comparison dimensions (execution freedom, governed information flow, identity separation).
\end{enumerate}

\textbf{Scope.} PES is a software design-and-implementation pattern for governed LLM agents. The claims are architectural and depend on a correctly implemented bridge (Section 4.1). Section 7 reports a mechanism check, a trace-isolation check, structural checks, and a five-model replication on the development/pilot implementation, plus a controlled cross-domain replication (regulated coding agents) that also bounds bridge overhead in two environments. Costs and applicability are in Section 4.7. Applicability under those conditions is not validation beyond the two environments.

\textbf{Structure.} Section 2 positions PES against related work. Section 3 gives the constructive argument for separation. Section 4 presents the pattern. Section 5 documents the case. Section 6 compares existing platforms and academic neighbors. Section 7 reports mechanism validation. Section 8 discusses applicability, limitations, and future work. Section 9 concludes.

\section{Background and Related Work}

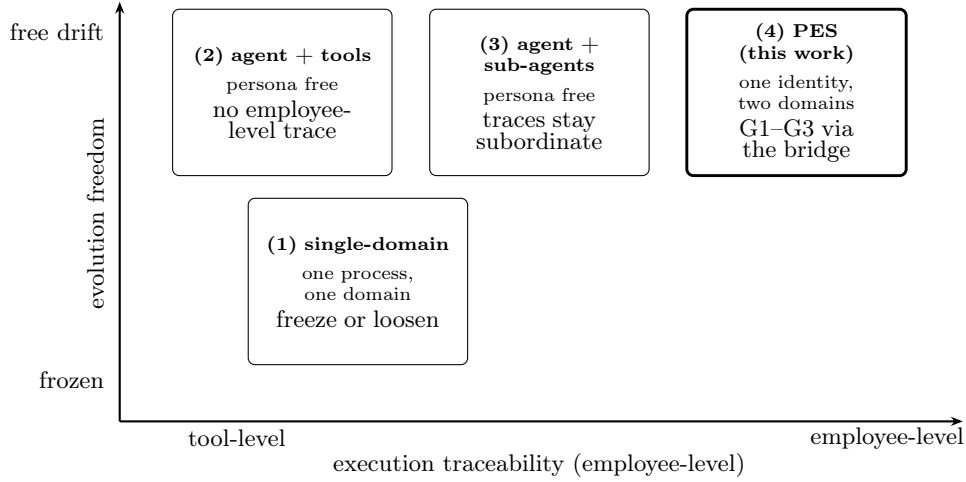
\begin{figure}[t]
\centering
\begin{tikzpicture}[
  font=\footnotesize,
  card/.style={
    draw, rounded corners=2.5pt, align=center,
    inner sep=4pt, text width=2.62cm, minimum height=2.2cm
  },
  slot/.style={card, fill=white},
  pes/.style={card, fill=white, line width=1.1pt},
  ax/.style={-{Stealth[length=1.6mm]}, thick},
  lab/.style={font=\scriptsize, align=center, inner sep=0.5pt}
]
\draw[ax] (0,0) -- (11.15,0);
\draw[ax] (0,0) -- (0,5.55);
\node[lab] at (5.55,-0.58) {execution traceability (employee-level)};
\node[lab, rotate=90] at (-0.32,2.75) {evolution freedom};

\node[lab] at (1.55,-0.22) {tool-level};
\node[lab] at (10.15,-0.22) {employee-level};
\node[lab, anchor=east] at (-0.18,0.55) {frozen};
\node[lab, anchor=east] at (-0.18,5.15) {free drift};

\node[slot] (p1) at (3.15,1.85) {%
  \textbf{(1) single-domain}\\[3pt]
  one process, one domain\\[2pt]
  {\scriptsize freeze or loosen}%
};
\node[slot] (p2) at (2.15,4.35) {%
  \textbf{(2) agent + tools}\\[3pt]
  persona free\\[2pt]
  {\scriptsize no employee-level trace}%
};
\node[slot] (p3) at (5.55,4.35) {%
  \textbf{(3) agent + sub-agents}\\[3pt]
  persona free\\[2pt]
  {\scriptsize traces stay subordinate}%
};
\node[pes] (p4) at (8.95,4.35) {%
  \textbf{(4) PES (this work)}\\[3pt]
  one identity, two domains\\[2pt]
  {\scriptsize G1--G3 via the bridge}%
};
\end{tikzpicture}
\caption{Four positions on the two axes that PES targets. Positions~(1)--(3) trade evolution freedom against employee-level execution traceability; PES occupies the remaining corner, paid for by the governed contract bridge (Section~4).}
\label{fig:spectrum}
\end{figure}

\subsection{The Agent Architecture Spectrum: Where PES Sits}

LLM agent architectures differ on two axes that PES targets: \textit{evolution freedom} (how freely the agent's persona can change without triggering re-validation) and \textit{execution traceability} (how strictly state-changing actions are gated and recorded). This cut is orthogonal to surveys that organize agents by construction modules (profile, memory, planning, action) \cite{wang-survey} and to SE-side work that treats agentic systems as architecture objects: observability patterns for reliable agentic software \cite{billiris}, and research agendas for generative AI in software engineering \cite{genai-agenda}; the IS community studies adjacent questions of agent robustness in security-critical settings \cite{radar}. Four positions cover the space (Figure~\ref{fig:spectrum}).

\textbf{Position (1): single-domain agents.} The default design: one process holds both the agent's identity (system prompt, instructions, persona) and its execution (tool calls, side effects). Chat assistants and most instruction-following agents (e.g., ChatGPT-class systems, ReAct \cite{react}) are here. The coupling is complete: persona and execution share a trust domain, so persona evolution and execution traceability cannot be tuned independently without reconstructing PES (Section 3.2).

\textbf{Position (2): agent + tools.} Execution is externalized as functions, tool calling \cite{toolformer}, function-calling loops \cite{openai-fc}, and framework ecosystems (LangChain, Dify, n8n). The persona is free (it lives in the agent), but the tools are stateless functions: they carry no employee-level identity, no cross-session memory, no audit persona. Trace is tool-level logs at best. The relationship is master--servant: the agent invokes, the tool obeys.

\textbf{Position (3): agent + sub-agents.} Execution gains autonomy, orchestrator-worker designs (AutoGen \cite{autogen}, CrewAI, OpenAI Swarm, and similar) delegate sub-tasks to sub-agents with their own loops. The persona is free and sub-agent traces exist, but the relationship remains master--servant, and the sub-agents are subordinate entities: there is no notion of one identity spanning trust domains, and sub-agent governance is typically uniform (all-or-nothing permissions).

\textbf{Position (4): PES (this work).} Persona and execution sit in different trust domains under one employee identity. Section 4 develops the pattern; Section 6 compares the positions.

\textbf{Terminology used throughout.} The two trust domains are a \textit{low-governance domain} (free evolution, light tracing) and a \textit{high-governance domain} (gated actions, full tracing). We call them the \textit{permissive domain} and the \textit{restrictive domain} after this point. The two faces are the \textit{expression surface} (the persona: where the agent is seen and edited) and the \textit{execution surface} (where it acts and is audited). The terms are defined here because they recur from Section 2 onward; Section 4 instantiates them concretely.

Positions (1)--(3) therefore trade the two axes: couple them, drop employee-level trace, or keep trace subordinate to the orchestrator. PES is the remaining corner, paid for by the bridge.

\subsection{Governance, Security, and Information Flow}

The pattern's trust-domain separation draws on three bodies of theory.

\textbf{Information-flow control.} Denning's lattice model \cite{denning} is the theoretical root of the one-way valve: security classes with permitted flow directions. PES instantiates that discipline at the organizational level rather than as a static lattice. The permissive and restrictive domains play the role of classes; the bridge channels are the permitted flows; enforcement is operational (ACL, approval, DLP, audit). The valve permits personal$\rightarrow$org only via approved promotion, a no-write-down analogue whose classes are organizational domains, not classification labels.

\textbf{The Bell-LaPadula model.} BLP \cite{blp} establishes the \textit{no-read-up / no-write-down} discipline for confidentiality: information may flow upward in classification but not downward. PES's one-way valve mirrors this: personal-space artifacts may enter the organization space (upward, via approved promotion) while organization-space data is \textit{by default forbidden} from flowing into the personal space (no downward read). The match is not exact (BLP governs subjects accessing objects by classification, whereas PES governs \textit{domains} with differing governance regimes) but the directionality discipline is the same, and Section 5 shows it was designed independently in the reference case (the one-way valve predates the PES framing).

\textbf{Policy enforcement: XACML PEP/PDP.} The eXtensible Access Control Markup Language \cite{xacml} separates the Policy Decision Point (PDP, where access decisions are made) from the Policy Enforcement Point (PEP, where they are applied). The analogy for PES is limited: the bridge's approval matrix is the decision point, and the domain boundary is the enforcement point. The difference is that XACML is a \textit{policy language and architecture for access control}, while PES is a \textit{pattern for agent identity and trust-domain placement}; the decision surface includes not just access but also data-egress grading and identity continuity. PES's bridge is a PEP-like enforcement point, but the policy it enforces is not an access-control rule set: it is the persona--execution relationship itself (what may cross, what must stay), which no attribute-based policy language currently expresses.

\subsection{Recent Work on Separation}

The most directly related recent work concerns separating aspects of LLM agent systems.

\textbf{Intent/execution separation (security angle).} Indirect prompt injection on tool-using applications \cite{greshake} (and its embodied-agent variants \cite{geng}; broader lifecycle threats, defenses, and evaluation for LLMs are surveyed in \cite{llmsec}) motivate this line. IsolateGPT \cite{isolategpt} isolates third-party app execution from the host system and from other apps (hub-and-spoke; anti-injection), the highest-impact execution-isolation architecture in this line, and Chahine \cite{chahine} separates \textit{intent} from \textit{execution} in a defense-in-depth architecture for multi-agent LLM systems, the nearest governance-side neighbor. Both cuts are security-motivated; IsolateGPT's object is the app execution surface, not a functional split inside one agent, and Chahine's is intent versus execution. PES's cut is governance/evolution-motivated and applies to one employee identity across trust domains: the isolated object is the persona (expression surface) versus audited execution, connected by a contract bridge with an audit anchor rather than a sandbox. The security framing has no first-class persona (expression surface). The two sit side by side: attack defense versus free drift under audit.

\textbf{Persona-conditioned errors (protective separation).} Recent empirical work documents \textit{persona poison} in LLM-controlled robots: persona prompts intended for dialogue altered safety-relevant navigation decisions in monolithic controllers, and the proposed mitigation is a separation-based architecture that keeps persona-conditioned prompts from reaching safety-relevant layers \cite{shaikh}. This is the closest \textit{empirical} confirmation that persona and execution interact in harmful ways, but the separation there is \textit{protective} (separate to prevent persona from contaminating safety-critical control), framed in a robotics/control setting, and explicitly \textit{non-adversarial} (the authors distinguish it from jailbreaking). The adversarial counterpart, an attacker tampering with an agent's persona/profile to overwrite identity beliefs, is studied as belief/profile poisoning (BPA-PP \cite{bpa}) and persona-modulation jailbreaks \cite{shah}. PES's separation is \textit{evolutionary} (separate so that persona may drift freely without contaminating audited execution) and applies to organizational agents whose execution is governed by approval matrices and audit chains; its layered drift (Section 4.3.4) additionally bounds adversarial profile tampering by pinning core identity and capability bindings in the restrictive domain. Protective split (persona must not corrupt control) and evolutionary split (persona must be free to change) both imply that persona and execution should not share a trust domain.

\textbf{Agent-to-workflow crystallization (temporal angle).} Progressive Crystallization \cite{pc} hardens exploration into deterministic, cheaper workflows over time. That is a lifecycle move, cost-motivated. PES is a placement at a given moment: persona and execution in different trust domains. This paper does not claim a conversion path between those forms.

\textbf{Governance infrastructure at the execution boundary.} Three contemporary systems build governance \textit{around} agent execution: Harness-MU \cite{harnessmu} argues that governance constraints should be deterministic runtime variables enforced by execution hooks rather than delegated to the LLM; the Organizational Control Layer \cite{ocl} separates proposal generation from environment-facing execution with policy interception and escalation; Agentao \cite{agentao} is a governed local-first runtime that separates model-generated action proposals from host-authorized execution through a host contract and a permission-mediated tool system, with state and execution traces as explicit abstractions. Its cut runs inside one governed runtime (proposals vs.\ authorized execution); PES's cut runs across two trust domains (the persona as expression surface vs.\ governed execution). These are the closest architectural neighbors to PES's bridge. These systems wrap execution (a gate, a runtime boundary). PES puts \textit{identity} across that boundary: the persona is deliberately not governed by the restrictive domain, and the bridge contract (summaries out / bodies stay / identity continuous) is about what may cross, not only about what may execute.

\textbf{Secure-execution cluster (2025--2026).} A 2025--2026 cluster attacks the \textit{same execution surface} as PES's bridge (approval, DLP, audit, least privilege) from a security angle: prompt injection, data exfiltration, over-privilege, and forged memory. They are complementary, not substitutes.

\begin{itemize}
\item \textbf{Planner-layer IFC.} Fides \cite{fides} is an agent planner that attaches confidentiality and integrity labels to messages, actions, and tool results, and deterministically enforces security policies before consequential actions (evaluated on AgentDojo \cite{agentdojo}). The overlap with PES is information-flow discipline; the objects differ. Fides labels \textit{data} to stop prompt injection and illicit flows inside one planner. PES places \textit{organizational trust domains} (permissive vs. restrictive) so a persona can drift without contaminating an audit ledger. Fides has no persona-as-identity, no dual-face employee, and no evolution goal (G1).
\item \textbf{Capability-enforced control/data flow.} CaMeL \cite{camel} extracts control and data flows from a trusted query and enforces capability metadata in a custom interpreter so untrusted retrieved data cannot hijack program flow (prompt-injection defense by construction). PES's ``capability binding'' is a different use of the word: a reference list of which SOPs an employee may initiate, not provenance tags on values. CaMeL secures a single agent against injection; it does not split operational identity across trust domains.
\item \textbf{Privilege policies at the tool-call interface.} Progent \cite{progent} encodes least privilege as symbolic rules over tool names and arguments, with SMT-checked policy updates (narrowing automatic; expansion requires approval: monotonic confinement). Adjacent to PES's deny/ask/allow approval matrix, but Progent shrinks an \textit{attack surface}; PES's matrix is \textit{organizational governance of cross-domain flow}. Progent does not model persona drift or a faceless execution surface.
\item \textbf{Runtime safety DSL.} AgentSpec \cite{agentspec} is a lightweight DSL of trigger / predicate / enforcement rules intercepting an agent's decision pipeline (stop, user inspection, self-reflection, safe alternative). Same family as Harness-MU/OCL: an external enforcement layer around execution. PES's contribution is not another rule language; it is the placement of identity across the boundary.
\item \textbf{Graded egress / data abstraction.} Firewalls \cite{firewalls} project communication onto task context: a Language Converter Firewall on inbound messages and a Data Abstraction Firewall (DAF) that abstracts outgoing personal data to a non-private, task-appropriate granularity rather than binary disclose-or-redact. \textbf{This is the closest neighbor to PES's ``summaries out, bodies stay'' plus data-egress grading.} The difference is motivation and identity: Firewalls is a privacy/security projection for agent-to-agent networks; PES is a governance/evolution contract between an employee's expression surface and its audited execution surface. DAF does not host a persona in a separate trust domain, and PES does not claim DAF's learned abstraction policies.
\item \textbf{Enterprise DLP + HITL.} SafeGPT \cite{safegpt} is a two-sided guardrail (input redaction, output moderation, human-in-the-loop policy feedback) wrapping enterprise LLM chat. Adjacent to the reference case's DLP middleware. SafeGPT wraps a \textit{chat model}; it has no dual-face employee identity, no SOP execution domain, and no free-drift goal.
\item \textbf{Tamper-evident execution ledger.} PoEM (Proof-of-Execution Memory) \cite{poem} keeps an HMAC-chained, append-only ledger of safety-critical actions that actually executed, so an agent may skip a safety step only if the ledger confirms it, defending against forged-reasoning memory (FARMA). Adjacent to PES's audit chain (G2). PoEM protects \textit{memory claims} about safety steps; PES's ledger is an \textit{organizational audit of governed execution}, anchored to core identity. PoEM does not separate persona from execution.
\end{itemize}

IFC, capabilities, privilege DSLs, runtime enforcement, graded egress, DLP, and hash-chained audit each have a 2025--2026 neighbor. \textbf{None of them places an agent's operational identity (persona) and its governed execution in different trust domains so that the persona may drift (G1) without contaminating the ledger (G2--G3).} That combination remains PES's claim (Section 6).

\textbf{Open-source engineering practice.} Multi-agent ecosystems are also being productized and studied on SE platforms \cite{carnemolla}. Open-source platforms themselves occupy adjacent pieces, not the combination. DeepSeek Harness (MIT) provides per-action sandboxing and an approval service but is single-user and does not model two trust domains; AgentScope \cite{agentscope} (Apache-2.0) offers a permission system (deny/ask/allow with rule learning) --- the reference case's approval matrix was developed with AgentScope as a stated reference --- but no domain separation or persona concept; cordum (BUSL-1.1) is an agent control plane with approval-gate workflows, but license-restricted and monitoring-layer rather than in-loop; StaffDeck (AGPL) is the closest open-source digital-employee platform, with state-machine SOPs and permission isolation, but without the dual-face persona/execution split. Section 6 tabulates these systematically.

\subsection{Change Isolation: The Theoretical Anchor}

The pattern's core mechanism, placing ``what may drift'' and ``what must be traced'' in different domains, is an instance of \textit{change isolation}, whose classical statement is Parnas's criteria for decomposing systems into modules \cite{parnas} and whose architectural-practice framing is standard \cite{bass}: modules should be decomposed so that a change to one aspect of the system does not require changes elsewhere. PES applies this principle to the LLM agent: the persona is the module that changes (drifts), the execution is the module that must not be perturbed, and the trust-domain boundary is the module boundary. The difference from Parnas's original setting is the subject of isolation: Parnas isolates \textit{implementation details} behind interfaces; PES isolates \textit{persona semantics} behind a governance contract; the thing that drifts is not a code detail but the agent's operational identity, whose drift in a single-domain design would force re-validation of the execution side. Section 4 returns to this distinction.

\subsection{Summary}

Existing architectures trade evolution freedom against execution traceability (Section 2.1). The theory PES uses is information-flow control, BLP directionality, and XACML decision/enforcement (Section 2.2), not a new security model. The 2025--2026 neighbors occupy intent/execution security, crystallization, external governance layers, and the secure-execution cluster (Section 2.3). None places the persona itself across trust domains with a governed contract bridge so that it may drift without contaminating audited execution.

\section{Design Goals: Why Persona and Execution Must Be Separable}

\subsection{The Tension}

Consider an agent whose actions are state-changing and regulated: they must be approved, recorded, and auditable. Operators still edit its instructions, tone, and skill bindings, sometimes weekly, sometimes daily. Both requirements are non-negotiable in the target setting.

If identity (persona) and action surface (execution) share a trust domain, that domain must be strict enough to make every execution auditable and permissive enough that persona edits are frictionless. Section 3.2 gives the argument its correct form under representational indistinguishability. Section 3.3 states the three goals; Section 3.4 notes the costs.

\subsection{Why Separation Is the Cheapest Resolution}

A naive version of the separation argument claims a \textit{logical} impossibility: that a single governance regime must treat persona and execution identically, so either persona evolution or execution traceability must fail. That version is unsound: a governance engine inside one domain can assign \textit{different rules by object type} (XACML does this per resource type; per-tool policies do this per tool). A single domain with differentiated policy could, in principle, give persona edits loose control and execution actions strict tracing. We therefore do \textbf{not} claim logical impossibility. The argument is about \textit{representation}, and it is specific to LLM agents.

\textbf{The representational indistinguishability premise.} In an LLM agent, the persona and the execution instructions are the \textit{same kind of thing}: natural-language context. Editing the persona is editing prompt text; changing execution semantics is also, at the level that matters to the model, changing prompt text. There is no compile-time or runtime signal that reliably separates ``this is a persona edit'' from ``this is an execution-semantics change,'' because both modify the same substrate. This is not hypothetical: persona content can be modulated to alter behavior \cite{shah}. In a single domain, the governance layer sees the same channel for both concerns and cannot distinguish them \textit{by construction}.

\textbf{The constructive claim.} Any mechanism that guarantees G1 and G3 (Section 3.3) within a single domain must introduce, somewhere, the following three elements:

\begin{enumerate}
\item \textbf{Typed change objects}: a way to declare, at edit time, ``this is a persona change'' vs. ``this is an execution change'';
\item \textbf{An external enforcement point}: a gate outside the LLM context (in-context gating can be overridden by injection) that applies different rules to the two kinds of change;
\item \textbf{A stable audit anchor}: a record that stays stable across persona edits, so identity remains continuous and traceability survives drift.
\end{enumerate}

But elements (1)--(3) \textit{are} PES: the trust-domain boundary is the external enforcement point; the contract bridge is the typed-change and audit-anchor mechanism. A ``single-domain'' solution is therefore not an alternative to PES: it is PES rebuilt inside the domain, at higher coupling cost. Each element has precedents in isolation (XACML-style gates \cite{xacml}, hash-chained execution ledgers \cite{poem}, typed objects in any structured-change system). The claim is constructive, not logical: jointly they reconstruct PES, and Section 6 shows that combination is currently absent. That is Parnas's form: change economics, not impossibility (Sections 2.4 and 4.1).

\textbf{Proposition 1 (constructive minimality).} Under the representational indistinguishability premise, any mechanism that delivers G1--G3 within a single trust domain must co-locate (i) typed change objects, (ii) an external enforcement point, and (iii) a stable audit anchor; PES is the cheapest construction of (i)--(iii) we found (as one governed crossing), not the only conceivable one.

\textit{Proof sketch.} (i)--(iii) are jointly sufficient by construction: typed objects give the domain a way to classify a change; the external gate applies the per-class rule outside the LLM context; the anchor keeps identity and trace stable across edits (Section 3.2). Necessity: without (i) the domain cannot distinguish persona edits from execution changes (premise); without (ii) in-context gating is overridable by injection (Section 2.3); without (iii) the audit record drifts with the persona (G2 violated). This is an architectural argument, not a formal impossibility theorem (Section 4.1).

\subsection{The Three Goals}

From the tension, three goals follow:

\textbf{G1: Free drift (evolvability).} Editing the persona (instructions, tone, self-presentation) must carry zero compliance cost: no re-validation of the execution side, no approval for the edit itself. The rationale is economic (every persona change pays the friction, repeatedly, in a continuously improved system) and organizational (the people who tune the persona --- operators, domain experts --- are not the people who certify executions).

\textbf{G2: Execution traceability (auditability).} Every state-changing action must be gated (approval where the policy requires), recorded in an audit chain, and \textit{unable to be retroactively contaminated} by persona changes. The execution side is the organization's record; its stability must not depend on how much the expression side drifts.

\textbf{G3: Decoupling.} Persona drift must not affect execution traceability. If the two concerns share a trust domain without typed change objects, an external gate, and a stable audit anchor, the auditability of past and future executions is entangled with the current state of the persona. That is a cost argument, not a logical impossibility (Section 3.2).

G3 is the load-bearing goal: G1 and G2 are individually satisfiable in a single domain (by permissive and restrictive governance respectively). Satisfying them \textit{together} without reconstructing typed change objects, an external gate, and a stable audit anchor (that is, without rebuilding PES) is what forces an architectural separation (Section 3.2).

\subsection{What Any Resolution Must Cost}

Separation is not free. Any resolution of the tension must accept three costs, which PES pays explicitly (Section 4.7 revisits them):

\begin{itemize}
\item \textbf{A bridge}: cross-domain communication is not free-form; it requires a governed channel with its own overhead (checks, approval lookups, audit writes).
\item \textbf{Two governance regimes}: operators must understand both the permissive and the restrictive domain (different policies, different audit postures, different operational footprints).
\item \textbf{Identity mapping}: one employee identity spanning two domains requires machinery to keep the two faces consistent (conversation$\leftrightarrow$run binding, continuity affordances).
\end{itemize}

The trade is: \textit{solve G1--G3 by construction of the domains, pay the bridge.} Section 4 presents that design; Section 7 reports the mechanism check on the development/pilot implementation.

\subsection{Summary}

Under representational indistinguishability, reconstructing typed change objects, an external gate, and a stable audit anchor inside one domain \textit{is} PES at higher coupling cost (Section 3.2). G1--G3 therefore favor an architectural separation, paid for by the bridge, dual regimes, and identity mapping. Section 4 presents that design.

\section{The PES Architecture Pattern}

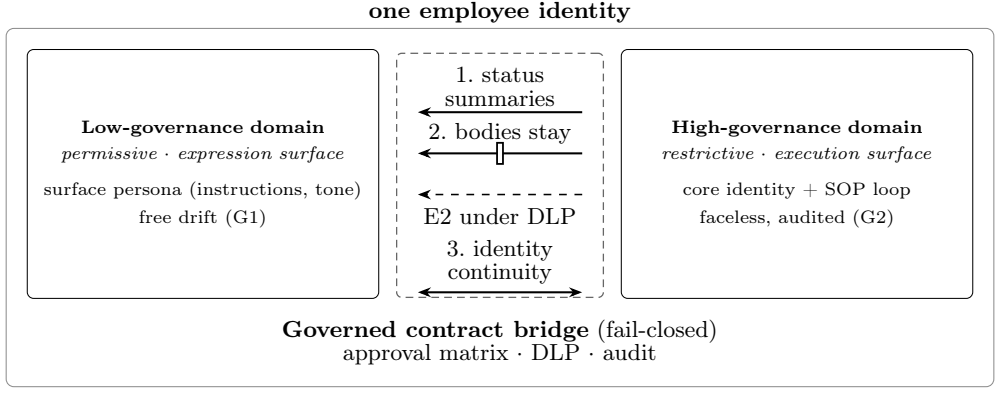
\begin{figure}[t]
\centering
\resizebox{\ifdim\width>\linewidth\linewidth\else\width\fi}{!}{%
\begin{tikzpicture}[
  font=\footnotesize,
  box/.style={
    draw, rounded corners=2.5pt, align=center,
    inner sep=5pt, text width=4.45cm, minimum height=3.4cm
  },
  permissive/.style={box, fill=white},
  restrictive/.style={box, fill=white},
  chan/.style={-{Stealth[length=1.8mm]}, thick, shorten <=15pt, shorten >=15pt},
  note/.style={font=\scriptsize, align=center, inner sep=0.3pt}
]
\node[permissive] (L) {
  \textbf{Low-governance domain}\\[2pt]
  \textit{permissive $\cdot$ expression surface}\\[6pt]
  surface persona (instructions, tone)\\[3pt]
  free drift (G1)
};
\node[restrictive, right=3.3cm of L] (S) {
  \textbf{High-governance domain}\\[2pt]
  \textit{restrictive $\cdot$ execution surface}\\[6pt]
  core identity + SOP loop\\[3pt]
  faceless, audited (G2)
};

\coordinate (L1) at ([yshift=24pt]L.east);
\coordinate (S1) at ([yshift=24pt]S.west);
\coordinate (L2) at ([yshift=8pt]L.east);
\coordinate (S2) at ([yshift=8pt]S.west);
\coordinate (L2e) at ([yshift=-8pt]L.east);
\coordinate (S2e) at ([yshift=-8pt]S.west);
\coordinate (L3) at ([yshift=-46pt]L.east);
\coordinate (S3) at ([yshift=-46pt]S.west);

% Channel 1: summaries flow back (high $\to$ low)
\draw[chan] (S1) -- (L1);
\node[note] (C1lab) at ($(S1)!0.5!(L1)+(0,10pt)$) {1.\ status\\summaries};

% Channel 2 baseline: bodies do not leave (closed gate on the path)
\coordinate (m2) at ($(S2)!0.5!(L2)$);
\draw[chan] (S2) -- (L2);
\node[fill=white, draw, line width=0.7pt, inner sep=0pt,
      minimum width=2.3pt, minimum height=8.2pt] at (m2) {};
\node[note] (C2lab) at ($(m2)+(0,8pt)$) {2.\ bodies stay};

% Graded exception: E2 body egress under DLP (dashed high $\to$ low)
\draw[chan, dashed, semithick] (S2e) -- (L2e);
\node[note] (E2lab) at ($(S2e)!0.5!(L2e)+(0,-8.5pt)$) {E2 under DLP};

% Channel 3: identity continuity is a round-trip
\draw[{Stealth[length=1.8mm]}-{Stealth[length=1.8mm]}, thick, shorten <=15pt, shorten >=15pt]
  (L3) -- (S3);
\node[note] (C3lab) at ($(L3)!0.5!(S3)+(0,12pt)$) {3.\ identity\\continuity};

% Fit the labels only (not the domain-edge docks) so side padding is
% around the text, and shorten keeps arrowheads inside the corridor.
\node[draw=black!60, densely dashed, line width=0.5pt, rounded corners=2pt,
      inner xsep=10pt, inner ysep=5pt,
      fit=(C1lab)(C2lab)(E2lab)(C3lab)] (BR) {};
\node[note, below=0.28cm of BR] (B) {%
  \textbf{Governed contract bridge} (fail-closed)\\
  approval matrix $\cdot$ DLP $\cdot$ audit
};

\begin{scope}[on background layer]
\node[draw=black!45, line width=0.4pt, rounded corners=3pt, fill=white,
      inner xsep=8pt, inner ysep=8pt,
      fit=(L)(S)(B)(BR),
      label={[font=\scriptsize\bfseries, inner sep=2pt]above:one employee identity}] {};
\end{scope}
\end{tikzpicture}%
}
\caption{The PES pattern: one employee identity spanning a low-governance and a high-governance trust domain, connected by a fail-closed contract bridge. Channel~2 is the baseline (barred); E2 knowledge-body egress under DLP masking is the graded exception (dashed) (Sections~4.4 and~7.3).}
\label{fig:architecture}
\end{figure}

\subsection{The Core Idea}

Throughout this paper, \textbf{ADR} refers to an \textit{Architecture Decision Record} \cite{nygard}, the reference case's governance artifact capturing each architectural decision with date, options considered, decision, and rationale (defined in Section 5.1).

\textbf{Persona--Execution Separation (PES)} is an architecture pattern \cite{gof, posa} in which an agent's \textit{expression surface} (persona: where it is seen, talked to, and edited) and its \textit{execution surface} (where it acts, and is audited) reside in \textbf{different trust domains}, connected by a \textbf{governed contract bridge}. Figure~\ref{fig:architecture} shows the pattern. The persona lives in a low-governance domain where it can evolve freely; execution lives in a high-governance domain where every action is traced. The bridge carries only what the design allows: status summaries flow back, data bodies stay in the restrictive domain except a graded data-loss-prevention (DLP) masked exception (Section 4.4), and identity stays continuous across the boundary.

Separation is how G1--G3 hold together. Section 3.2 states this as a constructive argument: a single domain could apply differentiated rules by object type, but under LLM representational indistinguishability any mechanism that actually delivers G1--G3 must re-introduce typed change objects, an external enforcement point, and a stable audit anchor. PES is the cheapest construction we found, not the only conceivable one. The claim is architectural and depends on the bridge (Section 4.4) being implemented as specified; we do not claim a formal guarantee. We use \textit{persona} for the agent's operational identity (instructions, tone, self-presentation), not a simulated character in dialogue-agent literature \cite{park}. A \textit{trust domain}, as used throughout this paper, is an operational governance boundary: inside it, one enforcement-mechanism set applies uniformly (one access-control list (ACL), one audit posture, one DLP policy); crossing it requires an explicit contract channel whose schema is validated at the boundary. This differs from a policy domain in XACML-style access control: the domain here is a scope over which enforcement machinery applies. The boundary itself is an architectural component, not a configuration.

\subsection{Design Goals: Drift and Traceability}

The pattern is motivated by the three goals of Section 3.3: G1 free drift, G2 execution traceability, and G3 decoupling. A single-domain architecture does not satisfy all three cheaply (Section 3.2).

\subsection{The Pattern: Two Faces, One Employee}

PES instantiates the goals through three design decisions, all visible in the reference case (Section 5).

\textbf{4.3.1 Dual-face model.} A single agent identity has two faces. The \textit{permissive-domain face} (frontstage) is where the agent is encountered: conversational interface, advisory answers, and the entry point for task initiation. The \textit{restrictive-domain face} (backstage) is where the agent executes: jobs driven by a standard operating procedure (SOP), approvals, audit records. Both faces belong to the \textit{same} employee identity; this is not a split into two agents, but one agent with two surfaces. The distinction is governance, not identity.

\textbf{4.3.2 Persona singly-homed; the restrictive domain is faceless.} The complete persona (instructions, tone, self-description) is hosted in exactly one place: the permissive-domain agent. The restrictive domain does not carry a conversational persona at all; its user-facing surface is the SOP form, the execution engine, and the ledger, not a chat personality. This decision eliminates the dual-source-of-truth problem: there is exactly one definition of ``who the agent is,'' and the execution side does not maintain a second copy that could drift out of sync. (The leftover persona field in the employee model was narrowed to task-internal broadcast tone, not a personality, precisely to preserve this single-homing; ADR-030 \S2, decision log 2026-08-17.) \textit{Alternative considered and rejected: hosting a read-only persona mirror in the restrictive domain, so the backstage could present a consistent personality in its forms.} This was rejected for the same reason projection was (Section 4.3.3): a second copy, however read-only in intent, is a second source of truth that must be kept in sync, and the restrictive domain's user surface (forms, engine, ledger) has no conversational use for a personality; it is a work surface, not a chat surface.

\textbf{4.3.3 Binding, not projection.} The permissive-domain agent relates to the restrictive-domain execution surface by \textit{capability binding}: a reference list of which SOPs and scenario packs the agent may initiate, carried as identifiers in governed tool calls. The binding references rather than copies. The SOP (its authority, versioning, and logic) stays entirely in the restrictive domain. An earlier design considered \textit{projecting} the persona into the restrictive domain (a copy, so the execution side could present a personality); this was rejected (ADR-030 \S3): projection would create two copies of the persona (reintroducing the drift problem PES exists to solve), while binding achieves the same user-visible capability with a single source of truth. \textit{Alternatives also considered: exposing the restrictive domain's full API to the permissive domain directly (rejected: it would widen the attack surface and defeat the checkpoint semantics of the bridge); and a free-form inter-agent channel (rejected: the work-order contract with schema validation is what keeps the crossing auditable, ADR-004).}

\textbf{4.3.4 Layered drift: core identity vs. surface persona.} G1's ``free drift'' is not unbounded. The persona is internally stratified into two layers with different drift permissions. The \textit{core identity} (the employee's name, roster entry, role boundaries, and the set of SOPs it is bound to) does not drift freely: it is the anchor to which audit records are attached, and changes to it are themselves governed (a role change is a promotion-like event, not an edit). The \textit{surface persona} (instructions, tone, self-presentation, skill-binding tuning) drifts freely with zero compliance cost. This stratification resolves the question ``if the persona drifts freely, what is the identity that the audit ledger attaches to?'': the ledger attaches to the core identity, which is stable by construction, while the surface persona is the layer permitted to evolve. The reference case embodies the split: a surface persona field (freely editable) versus the employee roster and its capability bindings (core, governed); ADR-030's narrowing of that field to task-internal broadcast tone is precisely a surface-layer adjustment, leaving the core untouched.

\subsection{The Governed Contract Bridge}

The two faces are connected by a contract bridge, not by free-form communication. The bridge carries three kinds of traffic, with asymmetric permissions. The channel set is closed by default: interactions outside the three channels are denied (fail-closed), not routed around; the bridge is the only crossing, and the crossing is the checkpoint (ADR-030 \S4). Note on scope: the three channels are \textit{data-plane} traffic (what information crosses between the faces). \textit{Control-plane} traffic --- the initiation of an execution (SOP selection, parameters) --- is not a fourth channel but part of the execution contract itself: it travels inside a work order whose schema is validated at the crossing (Section 4.3.3). The channel set is thus closed on the data plane and contracted on the control plane; there is no free-form path in either direction. The reference case's validation later legalized a fourth \textit{data-plane} class --- knowledge-body egress under DLP masking (E2 conditional egress, Section 7.3) --- as a graded exception inside the fail-closed design, not a free-form path. The three channels remain the design baseline; E2 is an evolution the checks exposed and then gated.

\begin{enumerate}
\item \textbf{Status summaries flow back.} The conversational face may see the \textit{progress} of restrictive-domain work: state, steps, responsible parties. This is what makes the frontstage useful as a command surface. But summaries are not bodies: the underlying artifacts (numbers, client details, files) do not leave the restrictive domain.
\item \textbf{Data bodies never leave (baseline).} The restrictive domain's data (workpaper figures, client records, documents) is not exported to the permissive domain as a free-form path. The contract defines a \textit{data-egress grading}: which fields count as ``summary-exportable'' versus ``body-restricted.'' In the reference case this grading is pinned in the shipped implementation (Section 7.3 S3): the E2 class above is the only legal exception, and only under DLP masking. The bridge is only as safe as that grading.
\item \textbf{Identity continuity.} A user moving from the frontstage to the backstage (e.g., deep-linking from a conversation to the ledger page that shows the work) must not experience a discontinuity: the backstage page renders a ``return to conversation'' affordance carrying the conversation ID, so round-trips do not break. The agent is one entity across the boundary, not two.
\end{enumerate}

The bridge's enforcement is layered: governed tool calls (Model Context Protocol (MCP) with ACL \cite{mcp}), an approval matrix (deny/ask/allow with a non-bypassable list), DLP checks on outbound content, and audit records for every crossing. The bridge is not a tunnel; it is a checkpoint. Approvals are keyed to the task: the matrix's call site is the task-mode authorization step, and the tool calls executed inside an SOP are gated by ACL, DLP, and the audit chain rather than re-approved per call (Section~7.5 reports both call sites). Compliance is enforced ``at the moment of entering the organization,'' not by surveilling the individual exploration that happens in the permissive domain.

\textbf{Threat surface.} PES is a governance/evolution pattern, not a security architecture; the security-angled separation of Chahine \cite{chahine} is complementary. Moving the persona into a low-governance domain enlarges the attack surface that the separation itself creates. Table~\ref{tab:threats} sketches three threats, the path, the defense on the bridge, and the residual. A full threat analysis is out of scope. The bridge defenses in Table~\ref{tab:threats} are additionally exercised against a declared set of attack classes; a bounded record of that check is in ESM~3 (Section~7.6).

\begin{table}[htbp]
\caption{Threat surface of the PES separation (sketch).}
\label{tab:threats}
\centering
\footnotesize
\setlength{\tabcolsep}{3pt}\renewcommand{\tabularxcolumn}[1]{m{#1}}\emergencystretch=1.5em
\begin{tabularx}{\textwidth}{>{\RaggedRight\arraybackslash}m{2.15cm}>{\RaggedRight\arraybackslash\hsize=0.90\hsize}X>{\RaggedRight\arraybackslash\hsize=1.25\hsize}X>{\RaggedRight\arraybackslash\hsize=0.85\hsize}X}
\toprule
Threat & Path & Defense on the bridge & Residual \\
\midrule
Prompt injection & Malicious frontstage message tries to initiate a harmful execution & Approval matrix (deny/\allowbreak ask/\allowbreak allow, non-bypassable); initiation is a work order & Social-engineering of an \textit{allowed} work order \\
Profile poisoning & Attacker with persona-edit access tampers with execution behavior & Layered drift: surface persona cannot change core identity or SOP bindings; DLP on egress & Surface persona still shapes \textit{phrasing} of allowed tools \\
Provider bypass & Permissive domain calls the model provider directly & Valve/\allowbreak ACL/\allowbreak DLP still govern domain-to-domain flow; bypass is invocation, not a second crossing & Model-invocation policy is outside PES \\
\bottomrule
\end{tabularx}
\end{table}

The correct antecedent for row 2 is not Shaikh \& Virkki's non-adversarial ``persona poison'' in robot navigation \cite{shaikh}, but belief/profile-poisoning (BPA-PP \cite{bpa}) and persona-modulation jailbreaks \cite{shah}. Shaikh \& Virkki remain a non-adversarial empirical precedent that separating persona from action selection reduces contamination; PES lifts that intuition to an organizational trust-domain architecture and adds the adversarial controls in Table~\ref{tab:threats}.

\subsection{Why This Is Not ``Agent-Calls-Tools''}

The permissive-domain agent can look like it merely \textit{calls} the restrictive domain as a tool (the standard agent-plus-tools architecture). Table~\ref{tab:agent-tools} states the contrast.

\begin{table}[htbp]
\caption{Agent-plus-tools vs. PES.}
\label{tab:agent-tools}
\centering
\footnotesize
\renewcommand{\tabularxcolumn}[1]{m{#1}}\emergencystretch=1.5em
\begin{tabularx}{\textwidth}{>{\RaggedRight\arraybackslash}m{2.15cm}>{\RaggedRight\arraybackslash\hsize=0.85\hsize}X>{\RaggedRight\arraybackslash\hsize=1.15\hsize}X}
\toprule
Dimension & Agent + Tools & PES \\
\midrule
\begin{tabular}{@{}l@{}}Nature of the\\execution side\end{tabular} & Stateless function (input $\rightarrow$ output) & \textbf{A complete digital employee}: state-machine execution, cross-session memory, audit trail, owned capability assets \\
Relationship & Master--servant (agent invokes tool) & \textbf{Two faces of one identity} (same employee, frontstage/\allowbreak backstage; persona singly-homed; identity continuous) \\
Contract & Function signature (I/\allowbreak O schema) & \textbf{Governance contract} (approval matrix, one-way valve, data-egress grading, audit chain) \\
Semantics & ``The agent calls a function'' & ``One employee operates across trust domains'' \\
\bottomrule
\end{tabularx}
\end{table}

The execution side is not a function; the relationship is not master--servant; the contract is not a schema. Section 6 shows no existing architecture provides that shift.

\subsection{Position in the Agent Architecture Spectrum}

PES is position (4) of the spectrum in Section 2.1 (Figure~\ref{fig:spectrum}): unlike single-domain agents, agent-plus-tools, and agent-plus-sub-agents, persona and execution sit in different trust domains, connected by the bridge of Section 4.4. Section 6 compares that placement with existing platforms and academic neighbors.

\subsection{Design Trade-offs and Applicability}

PES is not free. The separation costs:

\begin{itemize}
\item \textbf{Cross-domain communication overhead}: every execution initiation crosses the bridge (ACL check, approval lookup, DLP scan, audit write). For high-frequency trivial calls this is overhead that a co-located design does not pay.
\item \textbf{Two governance regimes to operate}: the permissive and restrictive domains have different policies, different audit postures, and different deployment footprints. Operators must understand both.
\item \textbf{Identity mapping complexity}: keeping one employee identity continuous across two domains requires the conversation$\leftrightarrow$run binding machinery (Section 5.4).
\end{itemize}

PES is indicated only when the following three conditions jointly hold: \textbf{multi-user or organizational deployment} (multiple operators share the agent, so persona changes are frequent and uncoordinated); \textbf{audit or compliance requirements on execution} (some state-changing actions must be traceable to a stable record, an obligation that regulatory frameworks increasingly make explicit \cite{westerstrand}); and \textbf{expected persona churn} (instructions, tone, or self-presentation are anticipated to change repeatedly). When any condition is absent, a single-domain design is defensible and cheaper. Bridge overhead is a \textit{cost} of the pattern, not a fourth applicability condition: even when all three hold, a team may still refuse the bridge if execution volume makes the crossing dominate. The pattern targets the regime in between: agents that must both evolve and be audited, and that can afford the crossing.

\subsection{Summary}

PES answers a specific failure of single-domain agent architecture: persona evolution coupled to execution traceability. Dual-face, persona singly-homed, and binding-not-projection, together with the bridge contract, are how G1--G3 are satisfied. Section 5 records how the pattern was converged upon in a development/pilot system.

\section{Case Study: FIA Workbench}

\begin{figure}[t]
\centering
\resizebox{\ifdim\width>\linewidth\linewidth\else\width\fi}{!}{%
\begin{tikzpicture}[
  font=\scriptsize,
  ev/.style={
    draw, rounded corners=2.5pt, inner sep=3.5pt,
    fill=white
  },
  arr/.style={-{Stealth[length=1.6mm]}, thick, shorten <=1.8pt, shorten >=1.8pt},
  gap/.style={font=\scriptsize, text=black!50, align=center, inner sep=0.6pt}
]
\def\evcard#1#2#3#4{%
  \parbox[c][2.42cm][c]{2.28cm}{\centering
    \textbf{#1}\\[1pt]
    {\scriptsize #2}\\[4pt]
    #3\\[5pt]
    {\scriptsize\textcolor{black!55}{#4}}%
  }%
}
\node[ev] (p1) {\evcard{P1}{19 Jul}{persona home}{none recorded}};
\node[ev, right=0.42cm of p1] (p2) {\evcard{P2}{19 Jul}{bind, don't copy}{not: per-employee copy}};
\node[ev, right=0.42cm of p2] (p3) {\evcard{P3}{19 Jul}{one-way valve}{not: unconstrained flow}};
\node[ev, right=1.25cm of p3] (p4) {\evcard{P4}{12 Aug}{promotion work order}{not: reuse delegation}};
\node[ev, right=0.78cm of p4] (p5) {\evcard{P5}{17 Aug}{dual-face crystallized}{not: mirror / project}};

\draw[arr] (p1) -- (p2);
\draw[arr] (p2) -- (p3);
\draw[arr] (p3) -- node[above, gap] {24\\days} (p4);
\draw[arr] (p4) -- node[above, gap] {5\\days} (p5);

\draw[{Bar[width=3.2pt]}-{Bar[width=3.2pt]}, black!40]
  ([yshift=-8pt]p1.south west) --
  node[below, gap, yshift=-0.4pt] {same day} ([yshift=-8pt]p3.south east);
\end{tikzpicture}%
}
\caption{Decision chain behind PES in the FIA Workbench pilot (2026-07-19 to 2026-08-17). P1--P3 occurred on the same day; later arrows mark calendar gaps. Grey text in each box is the rejected alternative (Table~\ref{tab:decisions}).}
\label{fig:timeline}
\end{figure}
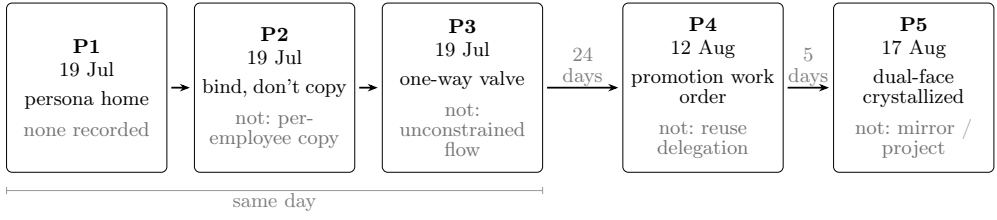

\subsection{Case Context}

The pattern is instantiated in the \textbf{FIA Workbench} (Financial-Industry AI Workbench), a digital-employee platform for financial institutions in a development/pilot deployment (a regulated, compliance-heavy industry where the tension motivating PES is not hypothetical). The workbench turns a financial firm's business processes, professional experience, and compliance requirements into persistent, auditable, evolvable digital employees that cover back-office operations (regulatory reporting, compliance filings, seal usage, capital calls), business support (industry research, due-diligence material analysis, IC memo drafting, post-investment data collection), and individual efficiency (scheduling, meeting minutes, personal knowledge bases).

\textbf{Naming.} The platform is referred to as FIA Workbench, a pseudonym. StaffDeck is named because it is an open-source (AGPL) reference. Model names in Section 7 are real. Internal component paths, tenant identifiers, and configuration keys are omitted.

\textbf{The IC memo task.} A central business task in this domain is the \textit{IC memo}, the Investment Committee memo that the investment team prepares before any investment decision. It is a structured decision document submitted to the firm's Investment Committee (partners, risk, and external experts) as the basis for deciding whether to invest, at what terms: it summarizes the target company, the diligence findings (financial/legal/commercial), the financial analysis (valuation, IRR/MOIC projections), the key risks with mitigations, the proposed terms, and a recommendation. The IC memo is the validation task because it is SOP-structured (a fixed template, so the restrictive domain can run it as a state machine), data-sensitive (diligence materials and figures exercise the bridge's ``bodies stay'' rule and the graded E2 exception, Section 4.4), and approval-gated (the draft must pass review before use). Concretely, the validation harness's V2 A/B pair (Section 7.4) executes the same fixed-input IC memo pipeline (retrieve, then generate the document) before and after a persona change and compares the execution records for contamination.

The IC memo validation uses one employee, not three roles. The sample-room agent is an investment-research assistant: its permissive-domain face carries the persona; its restrictive-domain face executes a versioned IC-memo SOP (retrieve, then generate) under approval. That is the dual-face of Section 4.3, bound rather than projected (Section 4.3.3). The same deployment also enforces dimension-ACL isolation and a DLP pipeline on the model gateway: the surfaces the bridge in Section 4.4 uses.

The regulated-finance setting was chosen because execution audit, approvals, and data-egress control are already first-class there, not as a convenience sample. Weaker-governance settings remain candidates under Section 4.7; they are not additional cases. The pilot is single-tenant: one firm's workbench, a one-month decision chain (2026-07-19 to 2026-08-17), and the sample-room SOP used by Section 7. We report it as a single-case study in the sense of Runeson and Höst \cite{runeson} and Yin \cite{yin}: a contemporary phenomenon in its real context, with internal validity from a complete decision record rather than statistical generalization. The tension is not manufactured, and the governance record is complete; every architectural decision is an ADR \cite{nygard} (date, options, decision, rationale); and translating such decisions into executable pipelines is itself an active architecture topic \cite{adrdevops}, while lesser items sit in a dated decision log.

\subsection{The Decision Chain Behind PES}

PES did not appear as a single design decision. It is the convergence of five decisions over one month, each directly about the persona--execution relationship (as opposed to the broader platform evolution reported elsewhere). Figure~\ref{fig:timeline} visualizes the chain; Table~\ref{tab:decisions} indexes it; the subsections narrate it.

\begin{table}[htbp]
\caption{Decision chain behind PES (FIA Workbench).}
\label{tab:decisions}
\centering
\footnotesize
\renewcommand{\tabularxcolumn}[1]{m{#1}}
\begin{tabularx}{\textwidth}{ll>{\RaggedRight\arraybackslash\hsize=1.10\hsize}Xll>{\RaggedRight\arraybackslash\hsize=0.90\hsize}X}
\toprule
ID & Date & Decision & Record & Role & Rejected \\
\midrule
P1 & 2026-07-19 & Persona gets a dedicated home, separate from execution & ADR-005 & Seed & None recorded \\
P2 & 2026-07-19 & Employees bind to shared capabilities by reference, not by copy & ADR-005 & \begin{tabular}{@{}l@{}}Binding\\prototype\end{tabular} & Copying skills per employee \\
P3 & 2026-07-19 & One-way valve: personal$\rightarrow$org via approval; reverse forbidden & Spec & Contract & Unconstrained bidirectional flow \\
P4 & 2026-08-12 & Promotion is a dedicated, approval-mandatory work order & ADR-014 & Channel & Reusing the ordinary delegation path \\
P5 & 2026-08-17 & Dual-face crystallized (singly-homed, faceless, bind-not-project) & ADR-030 & Crystallization & Persona mirror; projection; full API; free-form channel \\
\bottomrule
\end{tabularx}
\end{table}

\subsubsection{Seed: persona gets a home (P1, P2)}

The first ADR of the workbench (ADR-005, 2026-07-19) addressed two problems simultaneously; as the seed it predates the rejection-recording habit (Table~\ref{tab:decisions} records none for P1; P2, same day, already records one). Skills were employee-owned, so two employees reusing the same SOP had to copy it, and the copies drifted, defeating organization-level process standardization. And the employee model had no home for ``persona and duty instructions.''

The ADR's solution had a precedent: the reference implementation StaffDeck (an open-source digital-employee platform) modeled a digital employee's capability as \textit{bindings to four classes of organization-level resources} (SOP skills / general skills / knowledge bases / tools), with employees associated by binding relationships. The workbench adopted this: skills became workspace-level shared assets, and employees were associated with skills \textit{by reference}, including version pinning (pin a specific version, or follow the latest published). Meanwhile the employee model gained a dedicated persona field.

Two seeds for PES are planted here. First, the persona is given an explicit, dedicated storage location: it is now a named artifact with a schema, not an implicit property of the execution process. Second, the binding mechanism establishes the pattern that an employee's capability is a \textit{reference to organization-level assets}, not a copy: the same reference-not-copy semantics that PES later applies to the persona itself.

\subsubsection{Contract: flow direction is governed (P3)}

The same day, the product specification froze a ``one-way valve'' rule (Spec V1.0 \S1.3):

\begin{quote}
Personal-space artifacts may enter the organization space only via approval (``promotion''); organization-space data is \textit{by default forbidden} from flowing into the personal space; cross-project-group calls are constrained by dimension ACL.
\end{quote}

And \S3.3 pinned the mechanics: personal-space sessions, knowledge, and memory are owner-visible only and excluded from org retrieval and audit export (with necessary security metadata retained); the \textit{only} path for personal-space artifacts into the organization space is a work order $\rightarrow$ approval $\rightarrow$ copy-as-org-asset-with-source-recorded; org-space documents, run records, and employee assets may \textit{not} be referenced by personal-space runs; the runtime rejects them at validation time and records an access-violation audit event.

This is the contract layer of PES: the direction of flow between the two domains is not left to the agents; it is a governance rule enforced by the runtime, with audit. The one-way valve is what later becomes the bridge's asymmetry (status summaries out; data bodies stay except the graded E2 class in Section 4.4).

\subsubsection{Channel: the promotion work order (P4)}

On 2026-08-12, ADR-014 defined the semantics of the promotion work order, the operational realization of the one-way valve. A promotion work order is \textit{asset transfer} (a personal-space artifact, approved, is copied into the org space as a new asset with its source recorded), fundamentally different from a delegation work order, which is \textit{function call} semantics (the delegate executes and the output is validated against a contract schema and returned). The ADR explicitly rejected reusing delegation fields for promotion, because the two have different payload semantics and mixing them would fork the meaning of the output payload.

The channel is governed by construction: a promotion order can be created only in a pending-approval state; there is no approval-free path from personal to organization space. The approval chain follows the organizational policy (default single approver, escalable to four-eyes by risk level), and every status/approval transition writes an audit event.

This is the bridge's ``single checkpoint'' mechanism: cross-domain flow is a \textit{work order with mandatory human-plus-mechanism approval}, not a free channel. The mechanism predates the PES framing by five days; it was designed as the promotion path, and PES later recognizes it as the bridge's enforcement.

\subsubsection{Crystallization: the dual-face model (P5)}

On 2026-08-17, ADR-030 crystallized the pattern. Its eleven decisions include the three that constitute PES:

\begin{enumerate}
\item \textbf{Dual-face model}: one employee identity, two faces, a permissive-domain frontstage (conversation: advisory answers, task initiation, progress reporting) and a restrictive-domain backstage (SOP jobs, approvals, audit). The two faces belong to the \textit{same} employee; the distinction is governance, not identity.
\item \textbf{Persona singly-homed; the restrictive domain is faceless}: the complete persona lives in exactly one place, the permissive-domain agent. The restrictive domain's user surface is the SOP form, execution engine, and ledger, not a chat personality. The leftover persona field from P1 is narrowed to task-internal broadcast tone, preserving single-homing. \textit{Rejected alternative: a read-only persona mirror in the restrictive domain} (a second copy, however read-only, is a second source of truth).
\item \textbf{Binding, not projection}: the permissive-domain agent relates to the restrictive domain by capability binding (a reference list of which SOPs/scenario packs the agent may initiate, carried as identifiers in governed tool calls). \textit{Rejected alternative: projecting the persona into the restrictive domain} (projection creates two copies and reintroduces drift; binding achieves the same capability with one source of truth). \textit{Also rejected: exposing the restrictive domain's full API directly} (widens the attack surface, defeats the checkpoint), and \textit{a free-form inter-agent channel} (the work-order contract with schema validation is what keeps crossings auditable).
\end{enumerate}

The ADR also defined the bridge's three traffic channels with asymmetric permissions: status summaries flow back (progress/state/steps/parties), data bodies stay on the restrictive side as the baseline (workpaper figures, client records, documents; a \textit{data-egress grading} --- which fields are summary-exportable vs body-restricted --- was the remaining design pin, later frozen as a field-level grading including the E2 exception of Section 4.4), and identity stays continuous (deep-linking from conversation to the restrictive-domain ledger page, with a ``return to conversation'' affordance carrying the conversation ID). The channel set is closed by default: interactions outside the contracted channels are denied (fail-closed); the bridge is the only crossing, and the crossing is the checkpoint.

\subsection{Why PES Emerged as a Convergence}

The five decisions were not made under the banner of ``persona--execution separation.'' Each was made for its own immediate reason: data-model completeness (P1), process standardization (P2), compliance (P3), operational realization of the one-way valve (P4), and product-architecture consolidation after the platform merger (P5). Yet they converge on a coherent pattern. Three observations from the decision records support the claim that the convergence is genuine rather than post-hoc rationalization:

\textbf{1. The decisions are about different problems but the same seam.} P1 and P2 are about the \textit{data model} (where does persona live; how do employees relate to capability); P3 and P4 are about \textit{flow governance} (which direction, through what channel, with what approval); P5 is about the \textit{product architecture} (how the merged platform presents one employee across two surfaces). Different problem layers, same seam: the boundary between ``who the agent is'' and ``what the agent does.''

\textbf{2. The rejected alternatives are the pattern's negative image.} Every decision that could have collapsed the separation was explicitly considered and rejected: copying skills instead of binding (P2: drift), read-only persona mirror (P5: dual source of truth), persona projection (P5: drift again), full API exposure (P5: attack surface), free-form channel (P5: auditability). The rejection records show the design space was actually explored, not assumed.

\textbf{3. The pattern predates its name.} The one-way valve (P3) and the promotion channel (P4) exist in the spec and the ADRs as compliance mechanisms, five days before ADR-030 gave them the dual-face framing. PES is a recognition of what the system had already converged on, not a label imposed from outside.

\subsection{Current State and Open Items}

ADR-030 (approved 2026-08-17) is implemented: persona narrowed to broadcast tone, binding-not-projection via a capability-binding registry, data-egress grading, and the sample-room SOP used by Section 7. The remaining honest gaps:

\begin{itemize}
\item \textbf{Run$\leftrightarrow$conversation binding:} identity continuity (the ``return to conversation'' round-trip) still requires a run$\leftrightarrow$conversationId binding through the MCP bridge; it is a product-maturity item, not a prerequisite of the mechanism validation, and Section 7 does not exercise this channel.
\item \textbf{Not a production system:} the case is a development/pilot deployment. Section 7's measurements are from the test/pilot environment of that implementation, not from production operators or production-scale traffic.
\item \textbf{Mechanism validation executed:} Section 7 reports V1 as mechanism verification (R = 0.00 across five model configurations), V2 field-level invariance, S1--S4 before/after with dual-criterion reporting on S1a and S3, a cross-domain replication (Section 7.5).
\end{itemize}

\subsection{Case Summary}

The case provides what a single-case architecture paper needs: a decision chain, not a snapshot. Five decisions over one month, four with options-and-rejection records (P1, the seed, predates the recording habit), sit in a spec and ADRs that are internally auditable. The sequence seed $\rightarrow$ contract $\rightarrow$ channel $\rightarrow$ crystallization shows the pattern surviving different problem layers, rather than being assumed in one design pass.

\section{Comparison with Existing Architectures}

\subsection{Open-Source Ecosystem}

Table~\ref{tab:oss} compares PES's reference case against open-source platforms on three dimensions that operationalize the drift and flow requirements of G1--G3: \textbf{execution freedom} (whether persona-side change can proceed without re-validating execution), \textbf{information flow} (whether cross-domain flow is governed by a one-way valve / approval), and \textbf{identity separation} (whether persona and execution can reside in different trust domains). Audit integrity, the G2 record chain, is compared qualitatively in Section~2.3 and Table~\ref{tab:neighbors}. The comparison is based on public documentation and repositories accessed 2026-08-27.

\begin{table}[htbp]
\caption{Open-source agent platforms vs. the PES dimensions (G1--G3 operationalized).}
\label{tab:oss}
\centering
\footnotesize
\setlength{\tabcolsep}{3pt}\renewcommand{\tabularxcolumn}[1]{m{#1}}\emergencystretch=1.5em
\begin{tabularx}{\textwidth}{>{\RaggedRight\arraybackslash}m{2.05cm}ll>{\RaggedRight\arraybackslash\hsize=1.15\hsize}X>{\RaggedRight\arraybackslash\hsize=1.15\hsize}X>{\RaggedRight\arraybackslash\hsize=0.70\hsize}X}
\toprule
Project & License & Stars & Execution freedom & Information flow & Identity separation \\
\midrule
\begin{tabular}{@{}l@{}}DeepSeek\\Harness\end{tabular} & MIT & \textasciitilde{}199k & \partmark{} plugin tree + replaceable loop & \partmark{} per-action sandbox + approval (single-user) & \absnmark{} \\
Dify & Apache-2.0+ & \textasciitilde{}154k & \partmark{} app type fixed at creation & \absnmark{} & \absnmark{} \\
Coze Studio & Apache-2.0 & \textasciitilde{}22k & \partmark{} workflow or agent types & \absnmark{} & \absnmark{} \\
n8n & Fair-code & \textasciitilde{}203k & \partmark{} workflow + AI nodes & \absnmark{} & \absnmark{} \\
AgentScope & Apache-2.0 & \textasciitilde{}30k & \absnmark{} multi-agent orchestration & \partmark{} deny, ask, allow + rule learning & \absnmark{} \\
cordum & BUSL-1.1 & \textasciitilde{}0.5k & \absnmark{} & \partmark{} approval-gate + compliance firewall & \absnmark{} \\
StaffDeck & AGPL & \textasciitilde{}1.8k & \partmark{} SOP state machine + employees & \partmark{} isolation, publish, audit & \absnmark{} no dual-face \\
LibreChat & MIT & \textasciitilde{}42k & \absnmark{} chat surface & \absnmark{} & \absnmark{} carrier only \\
\begin{tabular}{@{}l@{}}\textbf{FIA}\\\textbf{Workbench}\end{tabular} & --- & --- & \textbf{\fullmark{} persona drift, no re-validation} & \textbf{\fullmark{} valve + matrix + DLP} & \textbf{\fullmark{} dual-face} \\
\bottomrule
\end{tabularx}
\vspace{4pt}
{{\footnotesize\textit{Note.} Legend: \fullmark{} full support · \partmark{} partial / adjacent · \absnmark{} absent. Apache-2.0+ is Dify's own license (Apache-2.0 plus multi-tenant and logo restrictions). Star counts from the GitHub API, accessed 2026-08-27.\par}}
\end{table}

\textbf{Reading the table.} No open-source project combines all three dimensions. DeepSeek Harness has a replaceable agent loop and per-action approval, but it is single-user and has no trust-domain model. Dify and Coze Studio fix the execution style when the app is created (workflow versus agent), which is not runtime-decoupled persona drift. AgentScope \cite{agentscope} has a permission matrix (the reference case's approval matrix was developed with AgentScope as a stated reference) but no domain separation or persona concept. cordum is a monitoring-layer control plane under a non-open license. StaffDeck is the closest digital-employee platform but does not split persona from execution. LibreChat is the kind of chat surface that can host a permissive-domain face; it is not itself a dual-face architecture. The three dimensions individually have adjacent implementations; their \textit{combination} in one architecture does not. The last row is this paper's reference case. Its three full-support marks are the claims Section 7 checks: execution freedom by V1, information flow by S2 together with the case record (Section 5), and identity separation by S1--S4 and V2.

\subsection{Academic Neighbors}

Table~\ref{tab:neighbors} compares the academic works closest to PES on four dimensions: \textbf{separation object} (what is separated), \textbf{motivation} (why), \textbf{mechanism} (how), and \textbf{case depth}. ``Pilot'' for this work means a development/pilot deployment with a recorded decision chain, not a production system.

\begin{table}[htbp]
\caption{Academic neighbors vs. PES.}
\label{tab:neighbors}
\centering
\footnotesize
\setlength{\tabcolsep}{3pt}\renewcommand{\tabularxcolumn}[1]{m{#1}}\emergencystretch=1.5em
\begin{tabularx}{\textwidth}{>{\RaggedRight\arraybackslash}m{2.15cm}>{\RaggedRight\arraybackslash}m{2.50cm}>{\RaggedRight\arraybackslash}m{1.80cm}>{\RaggedRight\arraybackslash}Xl}
\toprule
Work & Separation object & Motivation & Mechanism & Case depth \\
\midrule
Chahine \cite{chahine} & \begin{tabular}{@{}l@{}}intent vs.\\execution\end{tabular} & security & multi-agent security architecture & none reported \\
IsolateGPT \cite{isolategpt} & \begin{tabular}{@{}l@{}}app execution vs.\\system / apps\end{tabular} & security & hub-and-spoke isolation & prototype + benchmark \\
Shaikh \& Virkki \cite{shaikh} & \begin{tabular}{@{}l@{}}persona prompt vs.\\safety control\end{tabular} & protective & robot persona split & robotics preprint \\
Crystallization \cite{pc} & \begin{tabular}{@{}l@{}}exploration vs.\\workflow\end{tabular} & cost & temporal crystallization & IT ops (production) \\
Harness-MU \cite{harnessmu} & \begin{tabular}{@{}l@{}}constraints vs.\\behavior\end{tabular} & safety & hooks + runtime variables & multi-user harness \\
OCL \cite{ocl} & \begin{tabular}{@{}l@{}}proposal vs.\\env. execution\end{tabular} & governance & intercept + escalate & evaluation env. \\
Agentao \cite{agentao} & \begin{tabular}{@{}l@{}}proposals vs.\\host-authorized\\execution\end{tabular} & governance & host contract + mediated tools & local-first runtime \\
Fides \cite{fides} & \begin{tabular}{@{}l@{}}labeled data vs.\\actions\end{tabular} & security (IFC) & planner taint /\allowbreak  policy & AgentDojo \\
CaMeL \cite{camel} & \begin{tabular}{@{}l@{}}trusted flow vs.\\untrusted data\end{tabular} & security & capability interpreter & AgentDojo \\
Progent \cite{progent} & \begin{tabular}{@{}l@{}}allowed vs.\\extra tool calls\end{tabular} & least privilege & SMT privilege policies & AgentDojo /\allowbreak  ASB \\
AgentSpec \cite{agentspec} & \begin{tabular}{@{}l@{}}safe vs.\\unsafe actions\end{tabular} & runtime safety & trigger /\allowbreak  predicate DSL & code, embodied, AV \\
Firewalls \cite{firewalls} & \begin{tabular}{@{}l@{}}task data vs.\\overshare\end{tabular} & graded egress & converter + DAF & agentic-network bench \\
SafeGPT \cite{safegpt} & \begin{tabular}{@{}l@{}}sensitive IO vs.\\LLM\end{tabular} & enterprise DLP & two-sided guardrail & enterprise chat \\
PoEM \cite{poem} & \begin{tabular}{@{}l@{}}claimed vs.\\executed steps\end{tabular} & forged reasoning & HMAC execution ledger & LangChain agent \\
\textbf{PES} & \begin{tabular}{@{}l@{}}\textbf{persona vs.}\\\textbf{execution}\end{tabular} & \textbf{governance + evolution} & \textbf{dual domains + contract bridge} & \textbf{pilot (5 decisions)} \\
\bottomrule
\end{tabularx}
\vspace{4pt}
{{\footnotesize\textit{Note.} Abbreviations: AV, autonomous vehicles; DSL, domain-specific language; IFC, information-flow control; SMT, satisfiability modulo theories.\par}}
\end{table}

\textbf{Reading the table.} Two differences separate PES from every neighbor. First, the \textit{separation object}: existing work separates intent, governance, workflow, labeled data, control flow, tool privilege, safety rules, egress granularity, DLP boundaries, or memory claims from execution; IsolateGPT additionally isolates third-party app execution from the host system. Shaikh \& Virkki \cite{shaikh} is the closest \textit{object} neighbor (they separate a persona \textit{prompt} from safety-relevant control) but the setting is robotics, the motivation is protective and non-adversarial, and the execution side is not an organizational audit surface. \textbf{No neighbor separates an agent's operational identity from governed, audited execution so that the persona may drift.} Second, the \textit{motivation}: security (Chahine, IsolateGPT, Fides, CaMeL, Progent, Firewalls, PoEM), cost (crystallization), safety (Harness-MU/OCL/Agentao/AgentSpec, and Shaikh's protective split), and enterprise DLP (SafeGPT) are all served by moving things \textit{out of} or \textit{around} the execution surface. PES's motivation is different in kind; it moves the persona \textit{into} a low-governance domain precisely so that it can drift, while execution traceability is preserved. The closest \textit{mechanism} neighbor on the bridge is Firewalls' Data Abstraction Firewall (graded egress); the closest \textit{audit} neighbor is PoEM's hash-chained ledger. Both are cited as such in Section 2.3; neither hosts a dual-face identity. The closest temporal neighbor (crystallization) is orthogonal: it transforms along the time axis, whereas PES is a structural placement at a point in time.

\subsection{Positioning Summary}

Tables~\ref{tab:oss}--\ref{tab:neighbors} make the same point. G1--G3, operationalized as execution freedom under persona drift, one-way-valve information flow, and persona/execution identity separation, each have adjacent implementations in open-source platforms and academic neighbors (including the 2025--2026 secure-execution cluster of Table~\ref{tab:neighbors}), but no existing architecture provides all three in one design. PES is that combination by trust-domain placement, \textit{when the conditions of Section 4.7 hold}. The development/pilot case (Section 5) supplies what the academic neighbors lack: a decision chain showing the pattern was converged upon in engineering, not assumed.

\section{Validation: Mechanism Validation}

\subsection{What Needs Validation}

The pattern makes three claims a validation must address. V1 is \textbf{mechanism verification}; V2 is a runtime isolation check; V3 is an architectural necessity argument. V1 is not a competitive experiment.

\begin{itemize}
\item \textbf{V1 (mechanism verification of free drift):} persona changes incur \textit{zero re-validation cost} on the execution side. Formally: for every persona edit event, no execution-side re-validation event (re-approval, compliance re-review, execution-record invalidation) is triggered. \textbf{R = 0 is the designed outcome if the architecture is implemented correctly} (persona edits occur in the permissive domain and do not traverse the bridge; Section 4.3.2). A non-zero R would be an implementation defect, not a new empirical finding; the harness confirms the mechanism is present; it does not discover a cost advantage over a single-domain baseline.
\item \textbf{V2 (trace isolation):} execution records are \textit{not contaminated by persona drift}. Formally: audit records reference only the stable core identity (employee ID), never the surface persona; record contents are invariant under persona changes on hard-asserted semantic fields. Unlike V1, V2 exercises the running SOP and \textit{can} fail (model non-convergence, field-level contamination).
\item \textbf{V3 (necessity):} a single-domain architecture does not satisfy V1 and V2 \textit{cheaply}; doing so requires reconstructing PES inside the domain (Section 3.2). This is argued structurally; the validation probes the recovered pre-ADR-030 build empirically where possible (Section 7.4), and reports the finding as measured evidence.
\end{itemize}

\subsection{Validation Protocol}

\subsubsection{V1: Free-drift mechanism verification (executed 2026-08-21, measured)}

\textbf{Data sources} (system audit logs):
\begin{itemize}
\item Persona-change events: updates to the surface persona in the audit log.
\item Execution-side re-validation events: re-issued approvals, compliance re-reviews, invalidated execution records. Bridge denials of adversarial or injected requests (e.g., the L5 perturbations of Section 7.4) are entry denials (requests that never became approved executions), not re-validation events of an approved round; only re-approval, re-review, or record invalidation of an executed round count toward R.
\end{itemize}

\textbf{Metric:}
\begin{itemize}
\item R = (execution-side re-validation events) / (persona-change events)
\end{itemize}

\textbf{Pass criterion (mechanism, not discovery):} R = 0; no persona change triggers execution-side re-validation. Persona edits stay in the permissive domain and do not traverse the bridge (Section 4.3.2). Limits of this check are in Section~7.6.

\textbf{Control (V3):} recover a pre-ADR-030 build (2026-08-14) and ask whether persona edits enter governed execution on that build (Section 7.4). The probe is reported as measured evidence, not a reconstructed rate. We do not report a control-arm re-validation rate: the case record contains no written re-validation rule, so such a rate is not an observable on either arm. What the probe can show is whether the historical execution path consumed the persona.

\subsubsection{V2: Trace-isolation check (structural + data, executed 2026-08-21)}

\textbf{Structural check:}
\begin{itemize}
\item Inspect the audit schema: which identity fields do execution records reference? Expectation: employee ID (core identity) only, no surface-persona references.
\item Inspect the approval records: does approval re-validation depend on persona state? Expectation: no, approvals are keyed to work orders and core identity, not persona.
\end{itemize}

\textbf{Data check} (executed 2026-08-21; the A/B pair runs 33/34 around an L3 persona change):
\begin{itemize}
\item Verify content invariance on that pair. Metric: coupling between execution-record content and persona version = 0.
\end{itemize}

\subsubsection{V3: Necessity (argument + probe)}

\begin{itemize}
\item \textbf{Structural argument} (Section~3.2 states it in constructive form and rejects the logical-impossibility reading): under the representational-indistinguishability premise, a single-domain mechanism that delivers G1--G3 must re-introduce typed change objects, an external enforcement point, and a stable audit anchor; that is, PES rebuilt inside the domain at higher coupling cost. The claim is architectural, not statistical.
\item \textbf{Empirical probe:} a pre-ADR-030 build (2026-08-14) was recovered from version control; the probe is reported in Section 7.4 (\texttt{measured}, 2026-08-25). It does not supply a paired re-validation rate.
\end{itemize}

\subsection{Structural Checks (performed; pre-implementation baseline $\rightarrow$ post-implementation)}

Four architecture-constraint checks support V1--V2. They verify that the design \textit{excludes} the contamination paths by construction. \textbf{The checks were executed twice: a pre-implementation baseline (2026-08-21, static reviews of the codebase state \textit{before} ADR-030's dual-face model shipped) and a post-implementation re-run (2026-08-22, on the shipped implementation).} The baseline is deliberately reported in full (failures included): together with the post-implementation re-run it forms the before/after contrast the paper commits to. The post-implementation re-run re-enumerated every check from scratch (persona write points, governed-tool registry, edit-path tracing) rather than reusing baseline conclusions, and verified that post-baseline additions do not open new contamination paths. Table~\ref{tab:baseline} reports the baseline.

\begin{table}[htbp]
\caption{Structural checks, pre-implementation baseline.}
\label{tab:baseline}
\centering
\footnotesize
\renewcommand{\tabularxcolumn}[1]{m{#1}}\emergencystretch=1.5em
\begin{tabularx}{\textwidth}{ll>{\RaggedRight\arraybackslash}X}
\toprule
Check & Method & Baseline result (2026-08-21) \\
\midrule
S1a Single-homing & Edit-entry enumeration & \textbf{FAIL}: multiple write entries, including a chat-surface path (expected) \\
S1b No second copy & Second-carrier scan & \textbf{FAIL}: chat-surface store is a second writable persona (expected) \\
S2 One-way valve & Write-path + ACL inspection & \textbf{PASS} on writes; body-return tension counted in S3 \\
S3 Fail-closed channels & Registry vs. three channels & \textbf{FAIL (scope mismatch)}: retrieval returns bodies, not summaries \\
S4 Persona-edit path & Edit-chain tracing & \textbf{PASS} on reach; S1b still FAIL on replication \\
\bottomrule
\end{tabularx}
\end{table}

\textbf{Reading of the baseline.} S1a found multiple persona write entries, one reachable from the chat surface; S1b found that store independently writable and unsynced. S2's write path already held (promotion personal$\rightarrow$org only, mandatory approval, ACL, no org$\rightarrow$personal write); a read-side retrieval that returns document bodies is recorded under S3, not as an S2 miss. S4's PASS is on \textit{reach} (edits never hit restrictive-domain employee APIs) coexisting with S1b's FAIL on \textit{replication}. The two FAILs (S1a/S1b) and the S3 scope mismatch are the states ADR-030 was written to change: single-homing the persona (\S2), binding instead of projecting (\S3), and constraining the bridge to summaries (\S4). The two PASSes (S2/S4) show the \textit{contract} layer already in place; what is missing is the \textit{identity} layer. This baseline is the paper's ``before'' anchor.

\textbf{Post-implementation dual criteria (S1a and S3).} Two checks have a mechanical reading and a semantic reading. Reporting only one would either over-claim (PASS on semantics while the mechanical count is unchanged) or under-claim (FAIL on a spec-literal that the design itself evolved). Table~\ref{tab:dual} reports both readings; the PASS we rely on is the \textit{semantic / evolved} criterion: that is the pattern's claim.

\begin{table}[htbp]
\caption{S1a and S3: dual-criterion reporting (post-implementation).}
\label{tab:dual}
\centering
\footnotesize
\renewcommand{\tabularxcolumn}[1]{m{#1}}\emergencystretch=1.5em
\begin{tabularx}{\textwidth}{>{\RaggedRight\arraybackslash}m{2.45cm}>{\RaggedRight\arraybackslash\hsize=1.15\hsize}X>{\RaggedRight\arraybackslash\hsize=1.05\hsize}X>{\RaggedRight\arraybackslash\hsize=0.80\hsize}X}
\toprule
Check & Mechanical /\allowbreak\ spec-literal & Semantic /\allowbreak\ evolved & Paper verdict \\
\midrule
\textbf{S1a Single-homing} & Write-entry count unchanged (would stay FAIL) & Faceless execution: zero injection & \textbf{PASS} (semantic); count not hidden \\
\textbf{S3 Channel set} & Fourth class still returns bodies (would stay FAIL) & E2 under DLP; fail-closed remains & \textbf{PASS} (evolved); mismatch recorded \\
\bottomrule
\end{tabularx}
\end{table}

S1a's mechanical count did not drop: the employee model still has more than one write entry. The evolved claim is that the restrictive domain is faceless (persona narrowed to task-internal broadcast tone; reverse-assertion tested), not that a single write verb remains. S3's spec-literal still sees knowledge-retrieval as a fourth body-returning class, which would FAIL ADR-030 \S4 as first worded; egress grading records that class as E2 conditional body egress under DLP masking (three-plus-one), with fail-closed primitives intact. S1b, S2, and S4 do not need a dual reading: S1b flipped FAIL $\rightarrow$ PASS (the chat-surface store no longer carries a second persona; binding-not-projection implemented); S2 stayed PASS (one-way valve intact; post-baseline additions verified not to open an org$\rightarrow$personal write path); S4 stayed PASS (edit path stays in the permissive domain).

These are design-constraint checks, not experiments: they confirm the architecture removes the coupling structurally (the claim of Section 4), leaving the runtime side to V1--V2.

\subsection{Measured Results}

S1--S4 before/after is in Tables~\ref{tab:baseline}--\ref{tab:dual}. The harness ran on the development/pilot test deployment: 14 collected cases (self-checks, the five L1--L5 rounds, field-level A/B isolation, and audit-schema invariants), of which 12 pass, 2 skip by design, and 0 fail.

\textbf{V1 (mechanism verification).} Five perturbation rounds (L1 tone tweak $\rightarrow$ L5 adversarial ``skip approvals / rewrite audit'') produced \textbf{zero execution-side re-validation, R = 0/5 = 0.00 (measured=true)}, including L5 (limits in Section~7.6).

\textbf{V2.} Fixed-input A/B around an L3 persona change: state path, tool set, structural parameters, and approval chain identical; no surface-persona fingerprint. Free-text LLM-assembled arguments differed (recorded, not failed); an A/A probe (no persona change) showed the same class of divergence (model noise). Three A/B pairs had the A round repeat a retrieval while tool shape, structural parameters, and approval chains stayed identical (flash runs 37/38: three kb\_search calls versus one on B; kimi 58/59 and glm 60/61: three versus two); the SOP allows $\geq$1, no persona-attribution evidence appeared, and the divergence is recorded as model-behavior variance.

\textbf{Control arm: single-domain probe (executed 2026-08-25, measured).} No historical production single-domain configuration exists. A pre-ADR-030 development build (2026-08-14) was recovered from version control and probed on the same SOP and fixed input: five L3 persona perturbations, model deepseek-v4-flash (as in the V2 baseline). In that build a persona-injection path existed on the general execution surface, but the governed SOP path did not consume the persona. Result: \textbf{no persona influence on SOP execution}: tool sequence, state-transition path, and structural contract fields identical 5/5; directed L3 instructions aimed at the artifact were 0/3 honored. Free-text title and filename differed 5/5, at the same magnitude as the V2 A/A probe's model-autonomy noise (above). A positive control on the general execution surface (not the SOP path) honored the same class of instruction, so the injection pipeline itself works. An \textbf{explicit-injection probe} then wired the persona into the SOP path: execution changed in 2/2 rounds on both deepseek-v4-flash and v4-pro; a control injection of an unrelated instruction was likewise honored 2/2. Tool sequence and approvals stayed constant: once connected, injected text affects phrasing, not orchestration or governance. \textbf{Reading:} the pre-separation SOP surface was decoupled from the persona \textbf{by omission, not by construction}, an unwritten implementation fact that a later wiring change could reverse. PES makes that isolation a written, audited rule (ADR-030 \S2, Section 4.3.2): the restrictive domain is faceless \textit{by rule}, not by default. We report \textbf{no paired contrast}: a control-arm re-validation rate would be uninformative because the baseline's decoupling is not a governance choice it implements. The necessity claim is carried by Proposition 1 (Section 3.2). Protocols and anonymized round tables for the single-domain and explicit-injection probes are provided as Electronic Supplementary Material 1.

\textbf{Cross-model.} V1's zero replicated on five configurations (v4-flash, v4-pro, qwen3.8-max, kimi-k3, glm-5.3). V2 passed on four; qwen3.8-max did not converge on the sample-room SOP in any of 4 clean runs (looped retrieval until the round cap) --- recorded as a model-behavior finding, not concealed. This observation is the empirical entry to the advancement-timing insight of Section 8.4. Two gateway adaptation gaps (sampling-parameter constraint; URL join) were fixed same-day; provider capacity dominated re-run cost. Table~\ref{tab:ledger} reports the per-configuration attempt counts. Engineering spikes on the underlying runtime (channel reliability; retrieval strategy) are out of scope here.

\begin{table}[htbp]
\caption{Cross-model run ledger (attempt counts from model-call and run-event logs).}
\label{tab:ledger}
\centering
\footnotesize
\setlength{\tabcolsep}{4pt}\renewcommand{\tabularxcolumn}[1]{m{#1}}\emergencystretch=1.5em
\begin{tabularx}{\textwidth}{>{\RaggedRight\arraybackslash}m{2.55cm}>{\centering\arraybackslash}m{1.55cm}>{\centering\arraybackslash}m{1.85cm}>{\RaggedRight\arraybackslash}X>{\RaggedRight\arraybackslash}m{2.10cm}}
\toprule
Model & Attempts & Completed & \begin{tabular}{@{}l@{}}Excluded\\(reason)\end{tabular} & V2 \\
\midrule
deepseek-v4-flash & 13 & 13 & 0 & PASS \\
deepseek-v4-pro & 2 & 2 & 0 & PASS \\
kimi-k3 & 10 & 4 & 6 (provider capacity:\newline 400 temp×2, 429/\allowbreak timeout×4) & \begin{tabular}{@{}l@{}}PASS\\(4/\allowbreak 4 completed)\end{tabular} \\
glm-5.3 & 5 & 3 & 2 (gateway 404×1,\newline account rate-limit×1) & \begin{tabular}{@{}l@{}}PASS\\(3/\allowbreak 3 completed)\end{tabular} \\
qwen3.8-max & 8 & 0 & 4 (transport errors) +\newline 4 (round-cap abort, model behavior) & \begin{tabular}{@{}l@{}}unmeasurable\\(0/\allowbreak 4 clean)\end{tabular} \\
\bottomrule
\end{tabularx}
\vspace{4pt}
{{\footnotesize\textit{Note.} V1 is 0.00 (5/5 measured) on all five configurations. V1 is computed from audit-log events and does not require SOP run completion; qwen3.8-max's unmeasurable entry applies to V2 only.\par}}
\end{table}

V1 is measured on the harness's five perturbation rounds per configuration and is independent of run completion; V2 requires a completed A/B pair, so its verdict is scoped to completed runs. All excluded runs were provider/gateway faults except qwen3.8-max's four round-cap aborts, which are model behavior (Section 8.4). Configurations: deepseek-v4-flash and deepseek-v4-pro ran at temperature 0; glm-5.3, kimi-k3, and qwen3.8-max are reasoning models whose APIs force temperature 1. V1's assertions run on deterministic code paths independent of sampling; V2 requires model commitment, and the temperature asymmetry's bearing on the qwen3.8-max reading is discussed in Section 8.4.

S1--S4 after implementation (2026-08-22) is in Tables~\ref{tab:baseline}--\ref{tab:dual}: S1b flipped; S1a/S3 are reported on both the mechanical and the evolved reading. V3's control is the single-domain probe of Section 7.4 (executed, measured). Table~\ref{tab:results} summarizes the validation outcomes.

\begin{table}[htbp]
\caption{Validation results.}
\label{tab:results}
\centering
\footnotesize
\setlength{\tabcolsep}{4pt}\renewcommand{\tabularxcolumn}[1]{m{#1}}\emergencystretch=1.5em
\begin{tabularx}{\textwidth}{>{\RaggedRight\arraybackslash}m{1.85cm}>{\RaggedRight\arraybackslash}m{2.55cm}>{\RaggedRight\arraybackslash}X>{\RaggedRight\arraybackslash}m{3.35cm}}
\toprule
Item & Arm /\allowbreak\ round & Observation & Verdict \\
\midrule
V1 & PES, L1--L5 & $R_{\mathrm{PES}}$ = 0/\allowbreak 5 = 0.00 (measured) & Pass (mechanism intact) \\
V1 & Control (V3) & SOP path uninfluenced; by omission, not construction & No paired contrast (Proposition 1) \\
V2 & A/\allowbreak B (runs 33/\allowbreak 34) & State, tools, approvals identical; no persona fingerprint & Pass \\
V2 & Free-text params & Differ; A/\allowbreak A attributes to model noise & Recorded (expected) \\
Audit & Pre/\allowbreak post windows & Event types and required fields intact & Pass \\
V2 anomaly & Runs 37/\allowbreak 38, 58/\allowbreak 59, 60/\allowbreak 61 & A-side repeat retrieval; SOP-allowed; no persona link & Recorded \\
\bottomrule
\end{tabularx}
\end{table}

\subsection{Cross-Domain Replication: Regulated Coding Agents}

The financial case is a single deep case in one domain; transfer to a second domain is normally left to the applicability argument of Section 8.1. To probe it empirically we ran a controlled replication in the regulated-coding-agent shape of that section, on a standalone harness independent of the FIA Workbench codebase. Vaughan \cite{vaughan} applies PES to a coding agent (Codex CLI) in a public practitioner write-up and identifies the persona/execution coupling as its gap (execution-side writability of the persona file); the replication exercises that shape deliberately.

\textbf{Design (same V1 logic as Section 7.2).} The permissive-domain persona is a coding style (functional vs.\ object-oriented, comment density). The restrictive domain runs a code-generation SOP: a state machine (generate $\rightarrow$ syntax-check $\rightarrow$ report) whose steps are structural contracts, gated by the bridge (approval matrix, DLP, hash-chained audit) as in the financial case. Baseline rounds run under persona A (functional style); perturbation rounds under persona B (object-oriented, annotated). A positive control redefines the SOP (``skip the test step and report directly'') and must change execution: it distinguishes ``the persona does not cross'' from ``the executor is insensitive to every input.'' A fourth arm wires the persona into the SOP definition (harness version of the Section~7.4 explicit-injection probe): a persona-carried execution directive (``skip the test step'') must then change the skeleton, attributing the A/B invariance to the cut path rather than to an executor that ignores input. Arms C and D reach the same skeleton through different principals: C changes the SOP definition as its author would, while D shows the persona itself changes execution once wired. The two arms fix who may change what; PES removes the persona-to-step path by construction. Output-style fingerprinting (functional vs.\ OOP idioms in the generated module; probed per model) confirms the persona edit is consumed at the surface, so an unchanged skeleton is not a null response.

\textbf{Results.} Table~\ref{tab:minipilot} summarizes the outcome. The execution skeleton (tool sequence, gate verdicts, structural contract fields, audit chain) stayed invariant under persona perturbation in the isolated arms: \textbf{25/25} across five model configurations and four providers (deepseek-v4-flash, deepseek-v4-pro, glm-5.3, kimi-k3, qwen3.8-max; five rounds each). SOP redefinition changed execution in \textbf{25/25} (positive control), and the persona-wired arm changed it in \textbf{25/25} (injection control: the same persona instruction alters the skeleton once a persona-to-step path exists). Skeleton assertions run on deterministic code paths (under isolation the persona text reaches only the LLM system message), so the invariant is structural; the model-dependent evidence of persona consumption is the artifact-style shift, which followed the persona (functional $\rightarrow$ OOP) in every model probe. The audit hash chain was intact in every round. qwen3.8-max converged 5/5 here, unlike on the financial sample-room SOP (Sections 7.4 and 8.4). \textbf{Reading the counts:} the invariance arm is a construction check, not a discovery (under isolation the persona text cannot reach the steps by code structure, so 25/25 adds no information beyond the construction); the replication's evidence proper is the injection arm's attribution, the style fingerprint, the bridge bound (below), and qwen's convergence. The replication is a mechanism reproduction with the same boundary as V1: it shows the separation outcome holds in a second domain under controlled conditions; it is not a comparative evaluation against a single-domain coding-agent baseline.

\begin{table}[htbp]
\caption{Cross-domain replication in regulated coding agents (standalone in-memory harness; five model configurations across four providers; five rounds each; executed 2026-09-07).}
\label{tab:minipilot}
\centering
\footnotesize
\setlength{\tabcolsep}{4pt}\renewcommand{\tabularxcolumn}[1]{m{#1}}\emergencystretch=1.5em
\begin{tabularx}{\textwidth}{>{\RaggedRight\arraybackslash}m{2.4cm}>{\centering\arraybackslash}m{1.7cm}>{\centering\arraybackslash}m{2.15cm}>{\centering\arraybackslash}m{1.9cm}>{\centering\arraybackslash}m{1.3cm}>{\RaggedRight\arraybackslash}X}
\toprule
Model & \begin{tabular}{@{}c@{}}Persona\\$\rightarrow$ skeleton\\invariant\end{tabular} & \begin{tabular}{@{}c@{}}SOP redefinition\\$\rightarrow$ skeleton\\changed\end{tabular} & \begin{tabular}{@{}c@{}}Persona wired\\$\rightarrow$ skeleton\\changed\end{tabular} & Audit chain & \begin{tabular}{@{}l@{}}Artifact style\\shift\end{tabular} \\
\midrule
deepseek-v4-flash & 5/5 & 5/5 & 5/5 & intact & functional $\rightarrow$ OOP \\
deepseek-v4-pro & 5/5 & 5/5 & 5/5 & intact & functional $\rightarrow$ OOP \\
glm-5.3 & 5/5 & 5/5 & 5/5 & intact & functional $\rightarrow$ OOP \\
kimi-k3 & 5/5 & 5/5 & 5/5 & intact & functional $\rightarrow$ OOP \\
qwen3.8-max & 5/5 & 5/5 & 5/5 & intact & functional $\rightarrow$ OOP \\
\midrule
All (25 rounds) & \textbf{25/25} & \textbf{25/25} & \textbf{25/25} & intact & --- \\
\bottomrule
\end{tabularx}
\vspace{4pt}
{{\footnotesize\textit{Note.} Five rounds per model. The two deepseek configurations ran at temperature 0; glm-5.3, kimi-k3 and qwen3.8-max reject lower values (HTTP 400) and were served at 1. Round-level traces and per-step gate timings are archived in the supplementary material (mini-pilot).\par}}
\end{table}

\textbf{Bridge overhead across environments.} The bridge-cost question of Section 4.7 asks for a measured boundary rather than an argument. We instrumented the bridge in both environments with identical metric names (t\_acl, t\_dlp, t\_audit), identical reporting (ns-resolution timers; p50/p95/max; component sum per SOP round; share of end-to-end time), and a noop baseline arm in the product environment. Table~\ref{tab:bridge} reports both environments side by side.

In the replication harness (in-memory implementation), every in-SOP tool call passes a gate (tool-face ACL, content-level DLP, hash-chained audit write) with per-check medians of $\sim$1--2 $\mu$s (ACL), 5--7 $\mu$s (DLP) and 39--56 $\mu$s (audit write) across the five models; median per-model component sums per SOP round (three gates plus the entry approval; recomputed from the per-round times recorded in the shipped traces) are 0.18--0.22 ms, and the largest per-round share of end-to-end time observed was 0.008\% (end-to-end time is LLM-scale: per-model medians 4.6--29.2 s). Fifty concurrent bridge crossings through independent bridge instances did not degrade latency (entry-level p50 0.019 ms; shared-ledger contention is not exercised). In the FIA Workbench (the real product), the SOP form (LP quarterly-report SOP, ten rounds per arm, deepseek-v4-flash) measured t\_dlp p50 3.68 ms (n=130) and t\_audit p50 6.34 ms (n=20, database-backed write); t\_acl had no observation points: in SOP-form execution, tool calls inside the state machine do not traverse the approval matrix, whose single call site is the task-mode authorization step. A task-mode probe (ten governed rounds with a high-risk tool under an ask$\rightarrow$approve closed loop) measured t\_acl p50 1.27 ms (n=34), the other components of the same order. Component sums per round were 69.4 ms (SOP form) and 41.0 ms (task mode), i.e., 0.110\% and 0.079\% of end-to-end net time (62.9 s and 52.1 s, approval wait excluded); asynchronous approval wait is listed separately and excluded from the bridge (1.0--1.2 s per round).

\begin{table}[htbp]
\caption{Bridge overhead across environments (identical metric names; ns-resolution timers; p50 unless noted).}
\label{tab:bridge}
\centering
\footnotesize
\setlength{\tabcolsep}{4pt}\renewcommand{\tabularxcolumn}[1]{m{#1}}\emergencystretch=1.5em
\begin{tabularx}{\textwidth}{>{\RaggedRight\arraybackslash}m{3.4cm}>{\RaggedRight\arraybackslash}X>{\RaggedRight\arraybackslash}X>{\RaggedRight\arraybackslash}X}
\toprule
 & \multirow{2}{=}{\begin{tabular}{@{}l@{}}mini-pilot replication\\(concept, in-memory)\end{tabular}} & \multicolumn{2}{c}{FIA Workbench (product, deepseek-v4-flash)} \\
\cmidrule(lr){3-4}
 & & SOP form & Task mode \\
\midrule
t\_acl (per gate) & $\sim$0.001 ms & n=0 (structural) & 1.27 ms (n=34) \\
t\_dlp (per gate) & $\sim$0.005--0.007 ms & 3.68 ms (n=130) & 3.66 ms (n=150) \\
t\_audit (per write) & $\sim$0.04--0.06 ms & 6.34 ms (n=20) & 8.87 ms (n=11) \\
Median component sum per round & 0.18--0.22 ms & 69.4 ms & 41.0 ms \\
Share of end-to-end time & $\leq$0.01\% & 0.110\% & 0.079\% \\
End-to-end net time (median) & 4.6--29.2 s per model & 62.9 s & 52.1 s \\
\bottomrule
\end{tabularx}
\vspace{4pt}
{{\footnotesize\textit{Note.} Replication columns are a lower bound (per-gate in-memory checks) and product columns an upper bound (real DLP pipeline, database-backed audit writes); the gap is environment depth. t\_acl has no observation point in SOP-form execution (in-SOP tool calls do not cross the approval matrix, whose only call site is task-mode authorization), so its value comes from a ten-round probe (10/10). Per-round sums cover three gates plus the entry approval. The FIA per-round figures (69.4~ms / 41.0~ms) are the median of the recorded per-round totals, not the product of the per-gate p50s and $n$ per round; those two estimators need not agree.\par}}
\end{table}

\textbf{Reading.} Absolute costs differ as expected: the replication is a pure in-memory implementation (the mechanism's lower bound), while the product carries a real DLP pipeline and database-backed audit writes (the upper bound). The conclusion is directionally identical: bridge overhead is well under 0.2\% of end-to-end time (dominated by the model call by orders of magnitude), even in the product. This is a bounded mechanism-overhead measurement, not a performance benchmark; component sums are reported rather than a noop-subtracted delta because noop-baseline differences are unresolvable. It excludes asynchronous approval wait and one-time SOP registration (Section 4.7).

\subsection{Threats to Validity of the Validation}

Limits of the validation:

\begin{itemize}
\item \textbf{Two environments, asymmetric depth.} The financial measurements come from one development/pilot deployment whose decision chain (Section 5) supports internal validity. The replication (Section 7.5) runs on an independent standalone harness built for the experiment: a simplified single-task SOP, sandboxed file effects, no multi-tenant operation, no product surface, and no decision chain; it demonstrates the mechanism in a second domain under controlled conditions, not a second deep case. External validity is argued as applicability under Section 4.7 (Section 8.1), not as demonstrated generality.
\item \textbf{Model coverage.} V1's zero replicated on all five configurations: separation lives in the architecture layer, not in model weights. V2 held on four; qwen3.8-max is unmeasurable on the financial sample-room SOP because it did not follow that SOP's contract (Section 7.4 ledger), while it converged 5/5 on the replication's closed-form SOP (Section 7.5; attribution in Section 8.4). Excluded runs were provider/gateway faults except qwen's four round-cap aborts. Third-party endpoint capacity dominated re-run cost and may affect replications.
\item \textbf{Model availability.} Configurations were called through provider APIs in September 2026; the harness records the configured model name in every round, and the traces shipped as ESM~2 (Section 7.5) and ESM~3 pin the strings used here; where the probe recorded the served identifier it differed from the configured name. Provider-side availability is not durable: on 2026-09-10 the provider retired the V4~Flash weights and now routes \texttt{deepseek-v4-flash} to a successor model, and in the same notice reversed an earlier plan to retire \texttt{deepseek-v4-pro} on 2026-09-14 (which therefore keeps serving under the same name); a re-run under a retired name would exercise different weights, and provider statements about the rest are themselves provisional, with no signal on the caller's side. This does not narrow the claim: the claim is architectural (Proposition~1), 25/25 invariance held in all five configurations, and the remaining configurations are served by other providers and are unaffected by this retirement.
\item \textbf{Baseline is static, not dynamic.} The S1--S4 checks (Section 7.3) are code-level reviews, not runtime tests: they confirm the \textit{architecture} excludes contamination paths by construction, but they do not exercise the system. The runtime side is carried by the harness (V1/V2), which was executed on the shipped implementation (Section 7.4). The paper does not conflate the two evidence types: structural checks show the design \textit{can} exclude the paths; the harness shows the running system \textit{does} behave as designed.
\item \textbf{V1 and the control arm (one reading).} R = 0 verifies that no path from persona edits into the approval/audit machinery exists in this implementation (including under L5), not that PES is cheaper than a measured single-domain baseline. The single-domain probe (Section 7.4) found the pre-ADR-030 SOP path decoupled from the persona (no persona consumption; explicit-injection probe shows wiring persona in does change execution), so \textbf{no paired contrast is reported}; the necessity claim is carried by Proposition 1 (Section 3.2), not by a measurement.
\item \textbf{Measurement scope.} R and the coupling metric do not measure bridge overhead; the bridge-overhead table (Section 7.5) bounds mechanism cost in two environments but is not a performance benchmark, excludes approval-checkpoint human wait and one-time registration, and reports component sums because noop-baseline end-to-end differences are unresolvable under model-latency variance.
\item \textbf{LLM nondeterminism vs. persona contamination.} Tool arguments vary even with zero persona change. Structural fields are hard-asserted; free-text divergence is attributed via A/A controls. Single-occurrence anomalies have weak statistical attribution.
\item \textbf{Permissive-domain write path deviation.} Rate-limiting blocked the chat-surface edit API during the run; persona perturbations were applied by equivalent writes to the same permissive-domain persona store, with mandatory restore. The restrictive domain has no path that reads that store either way, so the observation surface is unaffected.
\item \textbf{Post-run instrument corrections.} Three measurement/instrument fixes were made to the shipped harness (ESM~2) after the September runs (audit-ledger projection, denied-path timing, DLP coverage of payload keys); the run artifacts predate them and are reported unchanged, and none of the fixes touches the compared quantities; the invariance assertions read step skeletons and structural fields, and the bridge figures are recomputed from the run-recorded component sums (Section~7.5).
\item \textbf{Observer effects.} The system is operated by the development team; the audit data is from the development/pilot test deployment, not from production operators. The \textit{interpretation} of what counts as a ``re-validation event'' was defined in the protocol before measurement (to avoid post-hoc selection).
\end{itemize}

\noindent\textbf{Adversarial check (bounded).} A bounded check of the Table~\ref{tab:threats} bridge defenses, with the design frozen before the rounds, is in ESM~3; it is not a security evaluation. \textbf{One defect was found on the product arm and fixed}: concurrent double submission of the same approval could be accepted twice, because the decision write was not atomic.

\subsection{Summary}

On the shipped implementation, \textbf{V1 found $R_{\mathrm{PES}}$ = 0.00 across L1--L5}; \textbf{V2 passed on hard-asserted fields}; the single-domain probe (Section 7.4) found no persona influence on SOP execution; no paired contrast is available (Section~7.6). Across five model configurations, V1 reproduced on all five and V2 on four; qwen3.8-max did not converge on the financial sample-room SOP (Section 8.4). The cross-domain replication (Section 7.5) reproduced the separation outcome in a second domain (regulated coding agents), with 25/25 persona-perturbation invariance under isolation, 25/25 SOP-redefinition sensitivity and 25/25 persona-wired penetration (injection control) across five models and four providers, and bounded bridge overhead at well under 0.2\% of end-to-end time in both the replication harness and the product. S1b flipped FAIL $\rightarrow$ PASS; S1a and S3 pass only on the evolved reading. 

V1 is model-independent (persona edits never enter the bridge). V2 is not: it requires the model to commit to the SOP's terminal action. That is why the zero reproduced and the isolation check did not.

\section{Discussion}

\subsection{Applicability: When Section 4.7 Holds}

The evidence base is dual-domain: a deep development/pilot case in financial institutions (Sections 5 and 7.1--7.4) and a controlled replication in a second domain, regulated coding agents (Section 7.5). Applicability follows the three conditions of Section 4.7: multi-user or organizational deployment, audit or compliance requirements on execution, and expected persona churn. The replication exercises the mechanism under those conditions in simplified form; it is not extra-case validation in the sense of statistical generalization, and neither the case nor the replication is a claim that the pattern is domain-free. Healthcare, public administration, and legal services illustrate the same tension; this paper validates none of them. When any condition is absent, a single-domain design is defensible and cheaper.

To make the conditions recognizable rather than abstract, three deployment shapes instantiate them:

\begin{itemize}
\item \textbf{Multi-tenant organizational assistants (shared SOPs, per-team personas).} The same execution SOPs (issue a ticket, update the CRM, run a report) serve different teams, each binding a different persona: a helpdesk agent that should sound patient, a sales agent that should sound assertive. Persona edits are frequent and per-tenant; the underlying actions must still be traceable to a stable employee/role identity. This is the closest shape to the reference case outside finance.
\item \textbf{Regulated coding agents (persona vs. workspace side-effects).} A coding agent's persona layer holds style and communication preferences; its execution layer writes files, runs tests, and may touch credentials. In a single domain, a system-prompt edit such as ``refactor more aggressively'' is textually indistinguishable from a permission change: the same representational indistinguishability that motivates PES (Section 3.2). Vaughan \cite{vaughan} applies PES to Codex CLI's building blocks in a practitioner write-up and identifies precisely this coupling as the gap (execution-side writability of the persona file). Section 7.5 reports a controlled replication of the mechanism claims in this shape (five models, four providers, 25/25) so this is the one shape that is validated rather than merely illustrated.
\item \textbf{High-sensitivity document workflows (health, legal, government).} A case-summary or intake assistant must adapt its persona to the interlocutor, while records (medical files, case files, identifiers) remain in the restrictive domain. PES's graded DLP exception (summaries may return, bodies stay) and its audit anchor map onto cross-domain confidentiality requirements.
\end{itemize}

These deployment shapes are illustrations, not additional validations: the evidence is the financial case (Sections 5 and 7.1--7.4) plus the coding-agent replication of Section 7.5. They matter because they show the five decisions of Section 5 (persona storage, binding, one-way valve, channel, identity model) are questions any Section 4.7 deployment must answer, and because one shape (regulated coding agents) now carries both a practitioner write-up applying PES to that shape and a controlled replication.

Those five decisions are not finance-specific mechanisms: where the persona lives; how employees relate to capability; which direction data may flow; what channel; what identity model. Section 6 shows no open-source platform answers all five. PES is stated (Sections 3--4) in trust-domain terms. The case shows the pattern is realizable \textit{under Section 4.7}; it does not show statistical generality, and it does not show that PES dominates alternatives. The reference case's SOP-driven tasks (report generation, compliance documentation, knowledge-based retrieval) are the same shape as a wider class of LLM-based process automation: a state machine governs \textit{what} happens, an agent model decides \textit{how}, and the output is a structured artifact that must survive audit. Operators still tune instructions frequently (G1) while the artifacts and the actions that produce them must remain traceable (G2, G3). In that setting the SOP is the execution contract, the persona is the tunable surface, and the bridge keeps the two governed --- a placement pattern for such automation, not a new analysis technique.

\subsection{Relationship to Established Theory}

PES does not claim new security theory; it instantiates existing theory in a new setting. Section 2.2 anchors the pattern to information-flow control (Denning \cite{denning}), directional confidentiality (Bell-LaPadula \cite{blp}), and decision/enforcement separation (XACML \cite{xacml}).

\textbf{PES is change isolation applied to the agent.} Parnas's module decomposition isolates change behind interfaces \cite{parnas}; PES isolates the persona behind a governance contract. The new element is the \textit{subject} of isolation: not a code detail but an agent's operational identity, whose drift in a single-domain design would cascade into re-validation of the execution side. This extends the classical principle to the LLM-agent setting, where the ``module'' that changes is a prompt, a persona, a behavior, not a function.

\textbf{PES is directional, like BLP, but the direction is governed by organization, not classification.} The one-way valve (personal$\rightarrow$org requires approval; org$\rightarrow$personal is forbidden) mirrors BLP's no-read-up/no-write-down \cite{blp}, but the classes are organizational domains, and the permitted direction is enforced by an approval matrix and audit rather than a lattice. The reference case designed the valve independently of BLP (Section 5: the one-way valve predates the PES framing); the convergence is a point of confidence, not of copying.

\subsection{Limitations}

This paper does not establish:

\begin{itemize}
\item \textbf{Case-depth asymmetry.} One deep development/pilot case (finance, one team's decision culture) plus one controlled replication (coding agents, standalone harness). The replication tests the mechanism claims and the bridge overhead in a second domain under simplified conditions; it is not a second deep case: no decision chain, no multi-tenant operation, no production surface. Re-running the replication inside a real coding-agent product (the Codex CLI shape Vaughan \cite{vaughan} mapped) is left to future work. Following Runeson and Höst \cite{runeson} and Yin \cite{yin}, the decision chain (Section 5) supports internal validity of the deep case; external validity is argued as applicability under Section 4.7 (Section 8.1), not as statistical generalization.
\item \textbf{No comparative evaluation.} Section 7 reports a mechanism check and a trace-isolation check, not a measured comparison against a single-domain alternative; the pre-ADR-030 probe (Section 7.4) found the baseline decoupled by omission, so \textbf{no paired contrast} is available. The claimed advantage (G1--G3 by design) is architectural, as formalized in Proposition 1 (Section 3.2).
\item \textbf{Not a security analysis.} The threat sketch (Section 4.4) answers the obvious questions but is not a threat model; PES is a governance/evolution pattern, and security-angled separation (Chahine \cite{chahine}) is complementary.
\item \textbf{Implementation maturity.} The reference case is a development/pilot deployment, not a production system. ADR-030's four rings (persona narrowing, binding-not-projection, egress grading, sample-room SOP) are implemented; Section 7's mechanism validation ran on that implementation. Remaining product-maturity items (Run$\leftrightarrow$conversation identity-continuity binding; operators at production scale) are outside this paper's evidence base. The pattern's decisions are frozen; its operational realization is young.
\end{itemize}

\subsection{Future Work}

\textbf{Persona versioning.} If the surface persona drifts freely, execution records should cite the persona version in effect, so an audit can replay ``this run was under persona v2.'' That identity question is still open.

\textbf{Model identity in execution records.} The same question applies to the model behind a governed execution: on 2026-09-10 the provider of two configurations used here retired the V4~Flash weights and began routing that model name to a successor, and in the same notice reversed an announced retirement of the other configuration's model name, with no change visible to the caller either way. A name-level pin is therefore not an audit anchor: a record that cites a model name does not fix what produced it. Closing that gap means recording the served model identifier per round, alongside the persona version above.

\textbf{Coordination and hierarchy.} Multi-employee collaboration (one persona initiating another's governed execution) and a chain of trust domains (individual $\rightarrow$ team $\rightarrow$ organization $\rightarrow$ regulator) would extend the same placement; the reference case only sketches them.

\textbf{Re-validation as a first-class artifact.} The validation found no written rule mapping a persona change to a mandatory re-validation obligation. The single-domain probe (Section 7.4) measured whether the historical execution path consumed the persona; it did not measure that missing rule. The gap remains: such obligations are implicit in discussions of single-domain governance and vanish under PES, but no system in the case articulates them. Making the trigger an explicit approval-matrix class would make a re-validation \textit{rate} measurable. The probe we ran cannot substitute for that rule.

\textbf{Advancement timing: an unresolved attribution.} The cross-model arm of Section 7.4 raised a deployment question: before adopting a model on a governed SOP, verify whether it commits to the contract's terminal action once evidence suffices, rather than re-evaluating indefinitely. The question came from one model-behavior anomaly --- qwen3.8-max did not converge on the financial sample-room SOP (0/4 clean runs; repeated retrieval until the round cap) while lighter models converged --- but the anomaly sits at the intersection of two confounds: task form (an open-ended synthesis task with a large decision space, versus the replication's closed-form SOP of Section 7.5) and temperature (reasoning models are served with temperature forced to 1). A follow-up experiment (n=8 per condition; censoring-aware: round-cap rate plus medians) resolved neither confound: model and temperature showed no statistically discernible effect on convergence rounds (medians 5.0--6.0, overlapping, with high within-condition variance), and the earlier readings (``model-specific over-evaluation,'' later ``temperature-dominated'') were both over-attributions on small samples. The honest status is a \textbf{null result}: the original observation stands, its cause is \textbf{unexplained}, and task complexity of the financial sample-room SOP is the leading candidate (the same model converged 5/5 on the replication's closed-form SOP), left as an open question. We therefore do not cite the anomaly as evidence of a capability ordering, and the indirect over-evaluation reading is withdrawn. The engineering default, verify advancement behavior on a governed SOP before adoption, survives as a caution, not a finding. The follow-on research is contract-side: a coverage criterion that forces the terminal action, and per-round remaining-budget pressure. The state machine hardens \textit{what} happens (retrieve, then generate) and not \textit{when} the agent may advance. Full token-level forensics are not part of the argument.

\section{Conclusion}

Conventional architectures freeze the persona to keep execution auditable, or loosen execution governance so the persona can change. PES puts the two surfaces in different trust domains and pays for G1--G3 with the bridge.

In the FIA Workbench pilot, five decisions between 2026-07-19 and 2026-08-17, four recording a rejected alternative (the seed, P1, predates the habit; Section 5). Section 6 finds no neighbor that jointly satisfies G1--G3. Section 7 reports $R_{\mathrm{PES}}$ = 0.00 and field-level invariance on V2, a cross-domain replication and bridge-overhead measurements (Section 7.5); how to read those numbers is in Section~7.6. The pattern is realizable under Section 4.7. It does not force a choice between evolvable agents and auditable ones.

\FloatBarrier
\bmhead{Supplementary information}
Electronic Supplementary Material~1 (ESM~1) holds the Section~7.4 probe protocols and anonymized round tables. Electronic Supplementary Material~2 (ESM~2) ships the Section~7.5 replication as a runnable harness (source code, per-round traces, key-free dry run). Electronic Supplementary Material~3 (ESM~3) holds the design record of a bounded adversarial check of the bridge, frozen before its rounds, together with its runnable driver and per-round logs. Structural-check enumerations (Tables~6--7) and the five-model run ledger (Table~8) stay in the manuscript.

\section*{Statements and Declarations}

\subsection*{Funding}
The author did not receive support from any organization for the submitted work.

\subsection*{Competing interests}
The author has no competing interests to declare that are relevant to the content of this article.

\subsection*{Ethics approval}
Not applicable. This article reports an architecture pattern and a single-case study of a software system. It did not involve human participants or animals.

\subsection*{Consent to participate}
Not applicable.

\subsection*{Consent for publication}
Not applicable.

\subsection*{Data availability}
The reference case is a single-tenant development/pilot deployment; tenant-identifying configuration, production data, and the deployment's architecture decision records cannot be shared, which falls under the journal's exception for material whose sharing would compromise confidentiality and legal requirements. Where material can be shared it is shipped as Electronic Supplementary Material: ESM~1 (Section~7.4 probe protocols and round tables), ESM~2 (replication harness source, per-round traces, and a key-free dry run), and ESM~3 (adversarial-check design record, driver, and per-round logs). Structural enumerations (Tables~6--7) and the five-model run ledger (Table~8) stay in the manuscript; the ADR dates, options, decisions, and rejected alternatives are reported in Section~5.

\subsection*{Materials availability}
Not applicable.

\subsection*{Code availability}
The FIA Workbench implementation is not publicly released. The Section~7.5 replication harness is provided as ESM~2 (standalone source code, per-round traces, key-free dry run), independent of the reference codebase. Section~7.4 probe protocols are in ESM~1 (Data availability). The bounded adversarial check's design record, driver, and per-round logs are in ESM~3.

\subsection*{Author contributions}
Yisen Xi is the sole author and is responsible for the study conception and design, the case analysis, the validation, and the writing of the manuscript.

\bibliography{pes}

\end{document}